\documentclass[onecolumn]{aa}  
\usepackage{url,cancel,graphics,latexsym,epsfig,graphicx,amssymb,lscape,color,txfonts,natbib}     

\renewcommand{\d}[0]{{\rm d}}
\newcommand{\e}[0]{{\rm e}}
\renewcommand{\i}[0]{{\rm i}}
\newcommand{\ave}[1]{\langle #1 \rangle}
\newcommand{\Ave}[1]{\Big\langle #1 \Big\rangle}
\newcommand{\Ref}[1]{(\ref{#1})}

\newcommand{\ch}[1]{#1}
\newcommand{\wtilde}[1]{\widetilde{#1}}

\begin{document}

\title{Towards an understanding of third-order galaxy-galaxy lensing}

\author{Patrick Simon, Peter Schneider \& Daniela K\"ubler}

\institute{Argelander-Institut f\"ur Astronomie, Universit\"at Bonn, Auf dem
  H\"ugel 71, 53121 Bonn, Germany\\
  \email{psimon@astro.uni-bonn.de}}

\date{Received \today}

\authorrunning{Simon et al.}
\titlerunning{Towards an understanding of G3L}

\abstract{Third-order galaxy-galaxy lensing (G3L) is a next generation
  galaxy-galaxy lensing technique that either measures the excess
  shear about lens pairs or the excess shear-shear correlations about
  lenses. From their definition it is clear that these statistics
  assess the three-point correlations between galaxy positions and
  projected matter density.}
{For future applications of these novel statistics, we aim at a more
  intuitive understanding of G3L to isolate the main features that
  possibly can be measured.}
{We construct a toy model (``isolated lens model''; ILM) for the
  distribution of galaxies and associated matter to determine the
  measured quantities of the two G3L correlation functions and
  traditional galaxy-galaxy lensing (GGL) in a simplified context. The
  ILM presumes single lens galaxies to be embedded inside arbitrary
  matter haloes that, however, are statistically independent
  (``isolated'') from any other halo or lens position. Clusters of
  galaxies and their common cluster matter haloes are a consequence of
  clustering smaller haloes. In particular, the average mass-to-galaxy
  number ratio of clusters of any size cannot change in the ILM.
}
{GGL and galaxy clustering alone cannot distinguish an ILM from any
  more complex scenario. The lens-lens-shear correlator in combination
  with second-order statistics enables us to detect deviations from a
  ILM, though. This can be quantified by a difference signal defined
  in the paper. We demonstrate with the ILM that this correlator picks
  up the excess matter distribution about galaxy pairs inside
  clusters, whereas pairs with lenses well separated in redshift only
  suppress the overall amplitude of the correlator. The amplitude
  suppression can be \ch{normalised}. The lens-shear-shear correlator
  is sensitive to variations among matter haloes. In principle, it
  could be devised to constrain the ellipticities of haloes, without
  the need for luminous tracers, or maybe even random halo
  substructure.}
{}

\keywords{Gravitational lensing:weak -- Galaxies: halos --
  (Cosmology:) large-scale structure of Universe}

\maketitle


\section{Introduction}

Gravitational lensing \citep[][for a recent
review]{2006glsw.conf..269S} has established itself as valuable tool
for cosmology to investigate the large-scale distribution of matter
and its relation to visible tracers such as galaxies. In the currently
favoured standard model of cosmology
\citep[e.g.,][]{pee93,2003moco.book.....D}, the major fraction of
matter is a non-baryonic cold dark matter component, i.e., matter with
non-relativistic velocities during the epoch of cosmic structure
formation. In this situation, gravitational lensing is an excellent
probe as it is sensitive to all matter as long as it interacts
gravitationally.

The main observable in lensing is the distortion of shapes of galaxy
images, in the weak lensing regime mainly ``shear'', by the
intervening inhomogeneous gravitational potential that is traversed by
light bundles from the galaxy. Over the course of the past decade,
applications of the gravitational lensing effect have come of age. To
name a few results \citep[][and references
therein]{2010RvMP...82..331B}, it was used to map the dark matter
distribution, to study the matter density profiles in galaxy clusters
and to determine their masses, to measure the relation between the
galaxy and dark matter distribution (the so-called galaxy bias), to
constrain the total matter density of the Universe and its fluctuation
power spectrum, and very recently to gather independent evidence for
the overall accelerated expansion of the cosmos
\citep{schrabback2010}.

Of particular interest for this paper is the so-called galaxy-galaxy
lensing technique where positions of foreground galaxies (``lenses'')
are correlated with the (weak lensing) shear on background galaxy
images (``sources''). Thereby, statistical information on the
projected matter distribution around lens galaxies can be
extracted. Since the first attempt by \citet{1984ApJ...281L..59T} and
the first detection \citep{1996ApJ...466..623B,1996MNRAS.282.1159G} of
this effect, galaxy-galaxy lensing nowadays is a widely applied robust
method to study the galaxy-matter connection
\citep{2000AJ....120.1198F, 2001astro.ph..8013M, 2002MNRAS.335..311G,
  hvg02, 2003MNRAS.346..994P, 2004ApJ...606...67H,
  2004MNRAS.355..129S, 2004AJ....127.2544S, 2005MNRAS.362.1451M,
  2005A&A...439..513K, 2006MNRAS.368..715M, 2006MNRAS.370.1008M,
  2007A&A...461..861S, 2007ApJ...669...21P, 2011arXiv1107.4093V}. The
traditional and hitherto mainly employed approach is to correlate the
position of one lens with the shear of one source galaxy (GGL
hereafter).

\citet{2005A&A...432..783S}, \ch{SW05 hereafter}, advanced the
traditional GGL by considering a new set of three-point correlation
functions that either involves two lenses and one source (``correlator
$\cal G$'') or two sources and one lens (``correlator $G_\pm$'').
Both correlators are tools to directly study higher-order correlations
between galaxies and the surrounding matter field.  In the literature,
this technique is termed $3^{\rm rd}$-order galaxy-galaxy lensing or
galaxy-galaxy-galaxy lensing.  There are alternative but
mathematically equivalent ways to express these statistics, e.g., the
aperture statistics $\ave{{\cal N}^2M_{\rm ap}}$ and $\ave{{\cal
    N}M^2_{\rm ap}}$ instead of the three-point correlation functions
$G_\pm$ and $\cal G$ \citep[SW05,][]{2008A&A...479..655S}. In
practical measurements usually the aperture statistics are preferred,
as they automatically remove unconnected $2^{\rm nd}$-order
contributions in estimators of the statistics and allow one to
separate E-modes from B-modes or parity modes, of which the latter two
cannot be generated by gravitational lensing as leading order
effect. In this paper, we focus on the E-modes in the correlators
$\cal G$ and $G_\pm$.

G3L has already been measured in contemporary lensing surveys, such as
the Red-Sequence Cluster Survey \citep{2008A&A...479..655S},
and will therefore presumably be routinely measured with ongoing
surveys such as
KiDS\footnote{\url{http://www.astro-wise.org/projects/KIDS/}},
Pan-STARRS\footnote{\url{http://www.cfht.hawaii.edu/Science/CFHLS/}},
DES\footnote{\url{http://www.darkenergysurvey.org}}, or in the future
surveys Euclid\footnote{\url{http://sci.esa.int/euclid}, see also
  \citet{2011arXiv1110.3193L}} and
LSST\footnote{\url{http://www.lsst.org/}}. 
The prospects of learning more on the galaxy-matter relation or new
observational tests for theoretical galaxy models
\citep[e.g.][]{2004ApJ...601....1W,2006MNRAS.370..645B,
  2007MNRAS.374..809D} with G3L are thus quite promising.

The main obstacle for exploiting the new G3L statistics is their
physical interpretation. It is clear from the definition that $\cal G$
quantifies the shear signal (or projected matter density) in excess of
purely randomly distributed lenses, picking up only signal from
clustered lens pairs \citep[e.g.,][]{2006MNRAS.367.1222J}, and that
$G_\pm$ is a two-point correlation function of shear associated with
matter physically close to lenses. It is unclear, however, what
physical information this translates to and what new feature may be
contained in G3L that may be missing or is degenerate in traditional
GGL. To elucidate these new statistics and to pave the way for new
applications of G3L, we conceive here a simplistic model for the
distribution of lenses and matter: the isolated lens model. Then G3L
is flashed out in the light of this model. For the definition of
quantities relevant for weak gravitational lensing, we refer the
reader to \citet{bas01}. The mathematical machinery of a halo model
expansion devised in the calculations is very similar to
\citet{1991ApJ...381..349S}, although used in a different physical
context.

The structure of the paper lays out as follows. Sect. \ref{sect:ggl}
introduces our model and derives the tangential shear about a lens,
the GGL signal, expected from this description. Sect. \ref{sect:g}
moves on to calculate the lens-lens-shear or $\cal G$ correlator for
this specific scenario. Sect. \ref{sect:gpm} does the same for the
lens-shear-shear or $G_\pm$ correlator. The final
Sect. \ref{sect:discuss} summarises the main conclusions drawn in the
preceding sections. \ch{In the following sections, we are introducing
  a number of symbols that are listed in the Table \ref{tab:symbols}
  for clarity.}

\section{The isolated lens model and galaxy-galaxy lensing}
\label{sect:ggl}

\begin{table*}
  \caption{\label{tab:symbols}\ch{List of symbols used in the paper and their
    meaning. The flag denotes a real number for ``R'', a complex
    number for ``C'', or vector of numbers for ``V''.}}
  \begin{center}
  \begin{tabular}{lllp{0.3cm}lll}
    \hline\\
    Symbol & Meaning & Flag & &
    Symbol & Meaning & Flag\\\\\hline\\\\
    $\vec{\theta}$ & position on the sky & C &&
    $P_\alpha(\alpha)$ & p.d.f. of internal halo parameters & R\\
    $\theta$ & modulus of $\vec{\theta}$ & R &&
    $\vec{\theta}^{\rm h}_i$ & centre position of $i$th halo & C\\
    $\vec{\theta}_{ij}$ & difference vector
    $\vec{\theta}_i-\vec{\theta}_j$ & C &&
    $\overline{\gamma}_{\rm h}(\vec{\theta})$ & mean Cartesian halo shear
    profile & C\\
    $\theta_{ij}$ & modulus of $\vec{\theta}_{ij}$ & R &&
    $\overline{\gamma}_{\rm t,h}(\vartheta)$ & mean tangential halo
    shear profile & R\\
    $\delta_{\rm D}^{(2)}(\vec{\theta})$ & 2D Dirac delta function & R &&
    $\overline{\gamma}_{\rm t}(\vartheta)$ & mean tangential shear
    (GGL) & R\\
    $\gamma_{\rm c}(\vec{\theta})$ & Cartesian shear at $\vec{\theta}$
    & C &&
    $\delta\gamma_{\rm h}(\vec{\theta};\vec{\alpha}_i)$ & fluctuation
    of $i$th halo shear profile about $\overline{\gamma}_{\rm
      h}(\vec{\theta})$ & C\\
    $\gamma(\vec{\theta};\varphi)$ & shear at $\vec{\theta}$ rotated by
    $\varphi$ & C &&
    $n_{\rm g}(\vec{\theta})$ & lens number density (on sky)& R\\
    $\gamma_{\rm t}(\vec{\theta})$ & tangential shear at
    $\vec{\theta}$ relative to direction $\vec{\theta}$ & R &&
    $\bar{n}_{\rm g}$ & mean lens number density (on sky)& R\\
    $\gamma_\times(\vec{\theta})$ & cross shear at
    $\vec{\theta}$ relative to direction $\vec{\theta}$ & R &&
    $\kappa_{\rm g}(\vec{\theta})$ & lens number density contrast (on sky)&
    R\\    
    $\gamma_{\rm h}(\vec{\theta};\vec{\alpha}_i)$ & halo shear profile
    of $i$th halo& C &&
    $\omega(\vartheta)$ & 2pt-clustering of lenses (on sky)& R\\
    $\vec{\alpha}_i$ & internal parameters of $i$th halo & V &&
    $\Omega(\vartheta_1,\vartheta_2,\vartheta_3)$ & 3pt-clustering of
    lenses (on sky)& R
  \end{tabular}
  \end{center}
\end{table*}

Here we lay out a simple model for the distribution of matter and
galaxies inside it. This model is founded on the assumption that
lenses are embedded inside a matter halo that generates the shear
profile $\gamma_{\rm h}(\vec{\theta};\vec{\alpha})$ acting upon a
background source, where $\vec{\theta}$ is the separation vector from
the centroid of the galaxy, and $\vec{\alpha}$ denotes a set of
intrinsic halo parameters that control the matter density profile of
the halo. Importantly, the intrinsic parameters are statistically
independent of the intrinsic halo parameters of any other halo or the
separations of other lenses. We hence coin lenses and their host
haloes in this scenario ``isolated''.  In this sense, this is a very
crude halo model representation \citep{2002PhR...372....1C} of the
lens and matter distribution, assuming for simplicity that every halo
is occupied by exactly one (lens) galaxy. Notice that matter which is
statistically independent of the lenses does not need to be accounted
for, as this would not contribute to a galaxy-matter cross-correlation
function, although it certainly would affect the noise in a
measurement.

In the following, 2D positions on the flat sky are, for convenience,
denoted by complex numbers $\vec{\theta}=\theta_1+\i\theta_2$ where
$\theta_1$ ($\theta_2$) is the position in direction of the
$x$($y$)-axis. By
$\theta=|\vec{\theta}|=\sqrt{\vec{\theta}\vec{\theta}^\ast}$ we denote
the modulus of $\vec{\theta}$. Likewise the Cartesian shear 2-spinor
$\gamma_{\rm c}=\gamma_1+\i\gamma_2$ is denoted as complex number. We
define the tangential, $\gamma_{\rm t}$, and cross, $\gamma_\times$,
shear of $\gamma_{\rm c}(\vec{\theta})$ relative to the origin by
\begin{equation}
  \label{eq:rotshear}
  \gamma(\vec{\theta};\varphi):=
  \gamma_{\rm t}(\vec{\theta})+\i\gamma_\times(\vec{\theta})=
  -\e^{-2\i\varphi}\gamma_{\rm c}(\vec{\theta})\;,
\end{equation}
where $\varphi$ is the polar angle of $\vec{\theta}$.

\subsection{Isolated lens model}

\begin{figure}
  \begin{center}
    \epsfig{file=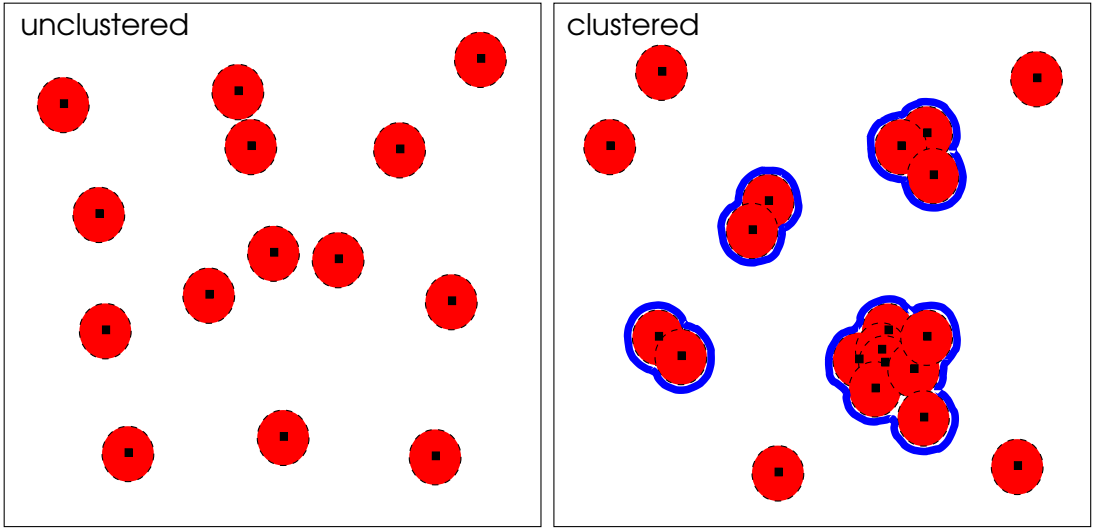,width=145mm,angle=0}
  \end{center}
  \caption{\label{fig:illu1} Illustration of the isolated lens
    model. Galaxies are depicted by black pixels, their matter haloes
    as red disks. For simplicity all haloes are identical in this
    visualisation. \emph{Left panel}: Lenses are distributed randomly
    on the sky. \emph{Right panel}: Lenses cluster to produce clumps
    with a common matter envelope, yet still described as sums of
    individual haloes. On the statistical level, this clumping is
    quantified by non-vanishing $2^{\rm nd}$-order and $3^{\rm
      3d}$-order correlation functions $\omega$ and $\Omega$,
    respectively.}
\end{figure}

Within the isolated lens model (ILM hereafter), the number density
distribution of lenses on the (flat) sky with area $A$ is
\begin{equation}
  \label{eq:tm1}
  n_{\rm g}(\vec{\theta})=
  \sum_{i=1}^{N_{\rm d}}\delta_{\rm
    D}^{(2)}(\vec{\theta}-\vec{\theta}^{\rm h}_i)\;,
\end{equation}
where $\delta_{\rm D}^{(2)}(\vec{\theta})$ is the Dirac delta
function, and the resulting shear field is
\begin{equation}
  \label{eq:tm2}
  \gamma_{\rm c}(\vec{\theta})=
  \sum_{i=1}^{N_{\rm d}}\gamma_{\rm h}(\vec{\theta}-\vec{\theta}^{\rm h}_i;\vec{\alpha}_i)\;,
\end{equation}
sticking a shear profile to every of the $N_{\rm d}$ lens position
$\vec{\theta}^{\rm h}_i$. The shear profile is directly related to the
projected matter density about the lens.  In the following we will use
the lens number density contrast
\begin{equation}
  \kappa_{\rm g}(\vec{\theta}):=\frac{n_{\rm
      g}(\vec{\theta})}{\overline{n}_{\rm g}}-1\;,
\end{equation}
where $\overline{n}_{\rm g}:=N_{\rm d}/A$ is the mean number density
of lenses within the area $A$. For the $2^{\rm nd}$-order angular
clustering correlation function of lenses on the sky
\citep[e.g.][]{peebles80}, we employ the function
\begin{equation}
  \omega\left(|\vec{\theta}_{12}|\right)=
  \Ave{\kappa_{\rm g}(\vec{\theta}_1)\kappa_{\rm g}(\vec{\theta}_2)}
\end{equation}
with \mbox{$\vec{\theta}_{ij}:=\vec{\theta}_i-\vec{\theta}_j$} being
the separation vector of two positions.  A value
\mbox{$\omega(\theta_{12})>0$} expresses an excess of galaxy pairs at
separation $\theta_{12}$ compared to a purely random distribution.

The ILM is a more general description than may appear at first sight:
Galaxy clumps with joint matter envelopes (``galaxy clusters'') are
not explicitly excluded, although every individual halo does host only
one lens. To form matter haloes of whole clusters, we can always stick
together and overlap matter haloes, changing the clustering
correlation functions in consequence.  The difference between the left
and right panel of Fig. \ref{fig:illu1} lies therefore in the choice
of the clustering correlation functions, which will enter the
following calculations. Hence, the ILM expands larger matter haloes as
sums of individual matter haloes of \emph{clustered} galaxies.
Crucially, however, the ILM is incapable to implement clumps of $N$
galaxies that contain on average more mass than $N$ isolated galaxies;
the mean matter-to-lens-number ratio has to be constant throughout. To
form a clump with a higher mass-to-lens number ratio would require to
increase the mass of all individual haloes simultaneously, i.e., to
change the internal halo parameters of all lenses inside the clump in
a similar fashion. This is not allowed, except by chance, however,
since internal parameters are statistically independent. A
generalisation of the ILM in this direction could be achieved in a
full-scale halo model with matter haloes hosting more than one
galaxy. Note that the luminosity of a lens can also be seen as an
internal parameter. Therefore, a constant mass-to-lens number ratio
plus statistical independence of internal parameters amounts also to a
constant mass-to-light ratio of all clumps in the model.

\subsection{Galaxy-galaxy lensing}
\label{sect:gglcalc}

\begin{figure*}
  \begin{center}
    \epsfig{file=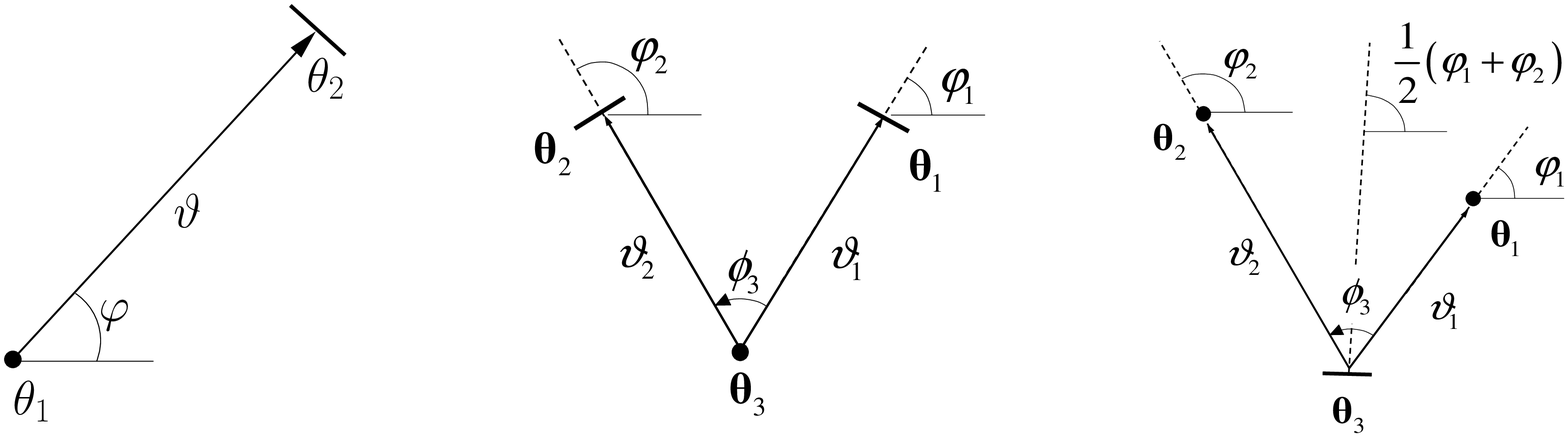,width=175mm,angle=0}
  \end{center}
  \caption{\label{fig:gglsketch} \emph{Left panel}: Illustration of
    the parametrisation of the GGL correlator $\overline{\gamma}_{\rm
      t}(\vartheta)$. The lens is located at $\vec{\theta}_1$, the
    source is at $\vec{\theta}_2$. \emph{Middle and right panel}:
    Illustration of the parametrisation of the G3L three-point
    correlators $\widetilde{G}_\pm(\vartheta_1,\vartheta_2,\phi_3)$
    (middle), and the galaxy-galaxy-shear correlation,
    $\widetilde{\cal G}(\vartheta_1,\vartheta_2,\phi_3)$ (right). The
    figures are copied from SW05.}
\end{figure*}

Before we embark on $3^{\rm rd}$-order statistics, we start with the
more familiar $2^{\rm nd}$-order GGL \citep{bas01}. These statistics
are the mean tangential shear $\overline{\gamma}_{\rm t}$ about a lens
at separation $\vartheta=|\vec{\theta}_2-\vec{\theta}_1|$
(Fig. \ref{fig:gglsketch}), defined by the correlator
\begin{equation}
  \label{eq:tanshear}
  \Ave{n_{\rm g}(\vec{\theta}_1)\gamma_{\rm
        c}(\vec{\theta}_2)}=
  -\e^{+2\i\varphi}\bar{n}_{\rm g}\left(
    \Ave{\kappa_{\rm g}(\vec{\theta}_1)\gamma(\vec{\theta}_2;\varphi)}
    +\underbrace{\Ave{\gamma(\vec{\theta}_2;\varphi)}}_{=0} \right)
  =-\e^{+2\i\varphi}\bar{n}_{\rm g}
  \,\overline{\gamma}_{\rm t}(\vartheta)\;.
\end{equation}
Owing to isotropy, the mean tangential shear $\overline{\gamma}_{\rm
  t}$ is only a function of separation $\vartheta$ and independent of
the polar angle $\varphi$. The bracket $\ave{\ldots}$ denotes the
ensemble average over lens number densities and shear
configurations. The underbraced term has to vanish due to the
statistical isotropy and homogeneity.

As lens number densities and shear configurations are expanded in
terms of haloes in the ILM, we consider the correlator as ensemble
average over all possible lens positions and internal halo parameters,
\begin{equation}
  \Ave{n_{\rm g}(\vec{\theta}_1)\gamma_{\rm c}(\vec{\theta}_2)}=
    \Ave{\sum_{i,j=1}^{N_{\rm d}}
      \delta_{\rm
        D}^{(2)}(\vec{\theta}_1-\vec{\theta}^{\rm h}_i)
      \gamma_{\rm h}(\vec{\theta}_2-\vec{\theta}^{\rm h}_j;\vec{\alpha}_j)}
  = 
  \underbrace{\sum_{i=1}^{N_{\rm d}}\Ave{\delta_{\rm
      D}^{(2)}(\vec{\theta}_1-\vec{\theta}^{\rm h}_i)
    \gamma_{\rm h}(\vec{\theta}_2-\vec{\theta}^{\rm
      h}_i;\vec{\alpha}_i)}_i}_{\rm one-halo}
  + 
  \underbrace{\sum_{i\ne j=1}^{N_{\rm d}}\Ave{\delta_{\rm
      D}^{(2)}(\vec{\theta}_1-\vec{\theta}^{\rm h}_i)
    \gamma_{\rm h}(\vec{\theta}_2-\vec{\theta}^{\rm
      h}_j;\vec{\alpha}_j)}_{i,j}}_{\rm two-halo}\;,
\end{equation}
which splits into two separate sums with ensemble averages over all
halo parameters of one halo (one-halo term)
\begin{equation}
  \Ave{\ldots}_i:=\frac{1}{A}\int\d^2\theta^{\rm h}_i\,\Ave{\ldots}_{\vec{\alpha}_i}
\end{equation}
or two haloes (two-halo term)
\begin{equation}
  \Ave{\ldots}_{i,j}:=
  \frac{1}{A^2}\int\d^2\theta^{\rm h}_i\d^2\theta^{\rm h}_j\,
  \left(1+\omega(|\vec{\theta}^{\rm h}_i-\vec{\theta}^{\rm h}_j|)\right)
  \Ave{\ldots}_{\vec{\alpha}_i,\vec{\alpha}_j}\;.
\end{equation}
The statistical independence of halo positions $\vec{\theta}^{\rm
  h}_i$ and internal halo parameters $\vec{\alpha}_i$ is explicitly
used here; $\ave{\ldots}_{\vec{\alpha}_i}$ and
$\ave{\ldots}_{\vec{\alpha}_i,\vec{\alpha}_j}$ are the averages over
internal halo parameters of a single halo or jointly for two haloes,
respectively. For the latter, we stress again that we will assume
statistical independence of $\vec{\alpha}_i$ and $\vec{\alpha}_j$.

In the following, we will need the average halo shear profile
\begin{equation}
  \label{eq:tanhalo}
  \overline{\gamma}_{\rm h}(\vec{\theta}):=
  \Ave{\gamma_{\rm h}(\vec{\theta};\vec{\alpha})}_{\vec{\alpha}}=
  \int\d\alpha\,P_\alpha(\vec{\alpha})
\gamma_{\rm h}(\vec{\theta};\vec{\alpha})=
-\e^{+2\i\varphi}\,\overline{\gamma}_{\rm t,h}(\theta)\;,
\end{equation}
$P_\alpha(\vec{\alpha})$ is the probability density distribution
function (p.d.f.) of the internal halo parameters $\vec{\alpha}$. The
tangential halo shear profile $\overline{\gamma}_{\rm t,h}$ is not to
be confused with $\overline{\gamma}_{\rm t}$ in Eq. \Ref{eq:tanshear}
that describes the total mean tangential shear about a lens including
contributions from the lens halo and haloes of clustering neighbouring
lenses.  Due to rotational symmetry, the average profile
$\overline{\gamma}_{\rm h}$ has a vanishing cross shear component, for
which reason we can express it in terms of the tangential halo shear
function $\overline{\gamma}_{\rm t,h}(\theta)$, which is only a
function of the separation $\theta$.

Utilising this definition, we arrive for the GGL correlator at
\begin{eqnarray}
  \Ave{n_{\rm g}(\vec{\theta}_1)\gamma_{\rm
        c}(\vec{\theta}_2)}
  &=&
  \sum_{i=1}^{N_{\rm d}}\frac{1}{A}
  \Ave{\gamma_{\rm
      h}(\vec{\theta}_{21};\vec{\alpha})}_{\vec{\alpha}}
  +
  \sum_{i\ne j=1}^{N_{\rm d}}\frac{1}{A^2}
  \int\d^2\theta^\prime\left[1+\omega(|\vec{\theta}^\prime-\vec{\theta}_{21}|)\right] \Ave{\gamma_{\rm
      h}(\vec{\theta}^\prime;\vec{\alpha})}_{\vec{\alpha}}\\
  &\approx&\nonumber
  \bar{n}_{\rm g}
  \overline{\gamma}_{\rm h}(\vec{\theta}_{21})+
  \bar{n}^2_{\rm g}
  \int\d^2\theta^\prime\omega(|\vec{\theta}^\prime-\vec{\theta}_{21}|)
  \overline{\gamma}_{\rm
    h}(\vec{\theta}^\prime)+
  \underbrace{\bar{n}_{\rm g}^2\int\d^2\theta^\prime\overline{\gamma}_{\rm
    h}(\vec{\theta}^\prime)}_{=0}\;.
\end{eqnarray}
The underbraced term must vanish due to radial symmetry.  Inside the
sums all terms become independent of individual lens positions
$\vec{\theta}^{\rm h}_i$ and halo parameters $\vec{\alpha}_i$ due to
the averaging. The last step assumes that the number of haloes is
large, i.e., \mbox{$N_{\rm d}\gg1$}, in particular \mbox{$N_{\rm
    d}(N_{\rm d}-1)\approx N_{\rm d}^2$}. This will also be assumed
for all following calculations.

Employing \Ref{eq:tanshear} and \Ref{eq:tanhalo}, we finally find
\begin{equation}
 \label{eq:ggl}
  \overline{\gamma}_{\rm t}(\vartheta)=
  \overline{\gamma}_{\rm t,h}(\vartheta)+
  \bar{n}_{\rm
    g}\int\!\!\d\theta^\prime\theta^\prime\d\varphi^\prime\,
  \e^{2\i(\varphi^\prime-\varphi)}
  \omega(\Psi)
  \,\overline{\gamma}_{\rm t,h}(\theta^\prime)
 =\overline{\gamma}_{\rm t,h}(\vartheta)+
  \bar{n}_{\rm
      g}\int_0^\infty\!\!\!\!\d\theta\,\theta\,\overline{\gamma}_{\rm
      t,h}(\theta)\!\!
    \int_0^{2\pi}\!\!\!\!\d\varphi
    \cos{(2\varphi)}
    \,\omega(\Psi)
\end{equation}
with the expression
\begin{equation}
  \label{eq:psi1}
  \Psi:=
  \sqrt{\theta^2+\vartheta^2-2\theta\vartheta\cos{\varphi}}\;.
\end{equation}
The last step in \Ref{eq:ggl} exploits that $\omega(\theta)$ has
vanishing imaginary part.  In the specific ILM description, the
average shear is expanded in terms of two components: The first term
in Eq. \Ref{eq:ggl} is the one-halo term, dominating at small
separations, whereas the second term is the two-halo term due to the
clustering of haloes.  Importantly, GGL is only sensitive to the
average lens halo $\overline{\gamma}_{\rm h}(\vec{\theta})$ but
insensitive to deviations of $\gamma_{\rm
  h}(\vec{\theta};\vec{\alpha})$ from $\overline{\gamma}_{\rm
  h}(\vec{\theta})$ in the actual halo population, which are
explicitly allowed within the ILM.

\section{Lens-lens-shear correlator}
\label{sect:g}

We now turn to the $3^{\rm rd}$-order galaxy-galaxy lensing
statistics, starting with the constellation of two lenses and one
source as depicted in the right panel of Fig. \ref{fig:gglsketch}. The
correlator considers a cross-correlation between lens number densities
at two positions $\vec{\theta}_1$ and $\vec{\theta}_2$ and the shear
at $\vec{\theta}_3$,
\begin{equation}
  \Ave{n_{\rm g}(\vec{\theta}_1)n_{\rm g}(\vec{\theta}_2)\gamma_{\rm
      c}(\vec{\theta}_3)}=
  -\bar{n}^2_{\rm g}\e^{+\i(\varphi_1+\varphi_2)}
  \,\wtilde{\cal G}(\vartheta_1,\vartheta_2,\phi_3)\;.
\end{equation}
Statistical homogeneity and isotropy implies that we can extract a
correlation function $\tilde{\cal G}$ from the correlator that is
solely a function of lens and source separations, $\vartheta_1$ and
$\vartheta_2$, and the opening angle $\phi_3$. This function is
split into an unconnected, $\wtilde{\cal G}_{\rm nc}$, and a connected
part
\begin{equation}
  {\cal G}(\vartheta_1,\vartheta_2,\phi_3)=
  \Ave{\kappa_{\rm g}(\vec{\theta}_1)\kappa_{\rm g}(\vec{\theta}_2)
    \gamma\left(\vec{\theta}_3;\frac{\varphi_1+\varphi_2}{2}\right)}
  =
  \wtilde{\cal G}(\vartheta_1,\vartheta_2,\phi_3)-
  \widetilde{\cal G}_{\rm nc}(\vartheta_1,\vartheta_2,\phi_3)\;.
\end{equation}
The connected part vanishes for unclustered lenses, while the
unconnected part can be shown to be generally (SW05)
\begin{equation}
  \label{eq:gunconnected}
  \widetilde{\cal G}_{\rm nc}(\vartheta_1,\vartheta_2,\phi_3):=
  \e^{-\i\phi_3}\overline{\gamma}_{\rm t}(\vartheta_1)+
  \e^{+\i\phi_3}\overline{\gamma}_{\rm t}(\vartheta_2)\;, 
\end{equation}
which is just the sum of the GGL shear profile around each
lens. Therefore, $\cal G$ encodes the shear in excess of what is
expected from unclustered lenses with the average shear profile
$\overline{\gamma}_{\rm t}$ around them, or: It quantifies the shear
signal about \emph{clustered} lens pairs. Note that the unconnected
terms do not contribute to the aperture statistics $\ave{{\cal
    N}^2M_{\rm ap}}$ and are thus not the primary quantity measured
with G3L.

\subsection{Derivation}

For $\cal G$ within the ILM, we need to evaluate the connected terms
of the correlator $\ave{n_{\rm g}(\vec{\theta}_1)n_{\rm
    g}(\vec{\theta}_2)\gamma_{\rm c}(\vec{\theta}_3)}$ with the model
specifics Eqs. \Ref{eq:tm1} and \Ref{eq:tm2},
\begin{eqnarray}
  \label{eq:tm3}
  \lefteqn{\Ave{n_{\rm g}(\vec{\theta}_1)n_{\rm g}(\vec{\theta}_2)
    \gamma_{\rm c}(\vec{\theta}_3)}=
  \Ave{\sum_{i,j,k=1}^{N_{\rm d}}
    \delta_{\rm D}^{(2)}(\vec{\theta}_1-\vec{\theta}^{\rm h}_i)
    \delta_{\rm D}^{(2)}(\vec{\theta}_2-\vec{\theta}^{\rm h}_j)
    \gamma_{\rm h}(\vec{\theta}_3-\vec{\theta}^{\rm h}_k;\vec{\alpha}_k)}}\\
&=&\nonumber
\underbrace{\sum_{i=1}^{N_{\rm d}}
\Ave{\delta_{\rm D}^{(2)}(\vec{\theta}_1-\vec{\theta}^{\rm h}_i)
  \delta_{\rm D}^{(2)}(\vec{\theta}_2-\vec{\theta}^{\rm h}_i)
  \gamma_{\rm h}(\vec{\theta}_3-\vec{\theta}^{\rm
    h}_i;\vec{\alpha}_i)}_i}_{\rm one-halo}
+
\underbrace{\sum_{i\ne j=1}^{N_{\rm d}}
   \Ave{\delta_{\rm D}^{(2)}(\vec{\theta}_1-\vec{\theta}^{\rm h}_i)
   \delta_{\rm D}^{(2)}(\vec{\theta}_2-\vec{\theta}^{\rm h}_j)
   \gamma_{\rm h}(\vec{\theta}_3-\vec{\theta}^{\rm
     h}_i;\vec{\alpha}_i)}_{i,j}+\texttt{(2~perm.)}}_{\rm two-halo}\\
&+&\nonumber
\underbrace{\sum_{i\ne j\ne k=1}^{N_{\rm d}}
    \Ave{\delta_{\rm D}^{(2)}(\vec{\theta}_1-\vec{\theta}^{\rm h}_i)
   \delta_{\rm D}^{(2)}(\vec{\theta}_2-\vec{\theta}^{\rm h}_j)
   \gamma_{\rm h}(\vec{\theta}_3-\vec{\theta}^{\rm
     h}_k;\vec{\alpha}_k)}_{i,j,k}}_{\rm three-halo}
\end{eqnarray}
where 
\begin{equation}
  \Ave{\ldots}_{i,j,k}:=
    \frac{1}{A^3}\int\d^2\theta^{\rm h}_i\d^2\theta^{\rm
      h}_j\d^2\theta^{\rm h}_k\,
    \big[
    \underbrace{1+\omega(\theta^{\rm h}_{ij})+
      \omega(\theta^{\rm h}_{ik})+
      \omega(\theta^{\rm h}_{jk})}_{\rm unconnected}+
    \Omega(\theta^{\rm h}_{ik},\theta^{\rm h}_{jk},\theta^{\rm h}_{ij})\big]
    \Ave{\ldots}_{\vec{\alpha}_i,\vec{\alpha}_j,\vec{\alpha}_k}
\end{equation}
is the ensemble average over three haloes (three-halo term). By
\begin{equation}
  \Omega(\theta_{13},\theta_{23},\theta_{12})=
  \Ave{\kappa_{\rm g}(\vec{\theta}_1)
    \kappa_{\rm g}(\vec{\theta}_2)\kappa_{\rm g}(\vec{\theta}_3)}
\end{equation}
we denote the (connected) $3^{\rm rd}$-order angular clustering
correlation function of the lenses that only depends on relative
galaxy separations. The sum \Ref{eq:tm3} hence decays into a one-halo
term (first sum), two-halo (next sum plus two identical sums apart
from permutations of the indices $i$ and $j$) and the three-halo term
(last sum). The one-halo term vanishes for
\mbox{$\vec{\theta}_1\ne\vec{\theta}_2$}, i.e., distinct lens
positions, owing to the Delta functions. For the connected terms, in
the halo correlators $\ave{\ldots}_{i,j}$ or $\ave{\ldots}_{i,j,k}$
only the summands with the leading order clustering correlation
functions are relevant, i.e., only the terms with $\Omega$ in the
correlator $\ave{\ldots}_{i,j,k}$, while terms with $\omega$ or 1 are
part of the unconnected terms (underbraced). They belong to the
unconnected part of the $3^{\rm rd}$-order lens clustering. Similarly,
in $\ave{\ldots}_{i,j}$ only terms associated with $\omega$ are of
relevance, the terms generated by $1$, the unconnected part of the
$2^{\rm nd}$-order lens clustering, will go into $\wtilde{\cal G}_{\rm
  nc}$.

By evaluation of all these ensemble averages one thereby obtains for
$\vec{\theta}_1\ne\vec{\theta}_2\ne\vec{\theta}_3$
\begin{equation}
  \label{eq:gtilde}
  {\cal G}(\vartheta_1,\vartheta_2,\phi_3)=
  {\cal G}_{\rm 2h}(\vartheta_1,\vartheta_2,\phi_3)+
  {\cal G}_{\rm 3h}(\vartheta_1,\vartheta_2,\phi_3)
\end{equation}
with the two-halo terms (we utilise the relation
$\overline{\gamma}_{\rm h}(-\vec{\theta})=\overline{\gamma}_{\rm
  h}(\vec{\theta})$ following from Eq. \ref{eq:tanhalo})
\begin{eqnarray}
  \nonumber
  {\cal G}_{\rm 2h}(\vartheta_1,\vartheta_2,\phi_3)&:=&
  -\e^{-\i(\varphi_1+\varphi_2)}\omega(|\vec{\theta}_1-\vec{\theta}_2|)
  \big(\overline{\gamma}_{\rm h}(\vec{\theta}_3-\vec{\theta}_1)+
    \overline{\gamma}_{\rm h}(\vec{\theta}_3-\vec{\theta}_2)\big)\\
   &=&\nonumber
    -\e^{+\i(\varphi_1+\varphi_2)}\omega(\vartheta_2)
    \big(
    -\e^{+2\i\varphi_1}\overline{\gamma}_{\rm t,h}(\vartheta_1)
    -\e^{+2\i\varphi_2}\overline{\gamma}_{\rm t,h}(\vartheta_2)
    \big)\\\nonumber\\
  &=&
  \label{eq:g2h}
  \omega(\vartheta_3)\big(
    \e^{-\i\phi_3}\overline{\gamma}_{\rm t,h}(\vartheta_1)+
    \e^{+\i\phi_3}\overline{\gamma}_{\rm t,h}(\vartheta_2)\big)\;,
\end{eqnarray}
the lens-lens separation
\begin{equation}
  \vartheta_3=
  \sqrt{\vartheta_1^2+\vartheta_2^2-2\vartheta_1\vartheta_2\cos{\phi_3}}
\end{equation}
and the three-halo term
\begin{eqnarray}
  \nonumber
  {\cal G}_{\rm 3h}(\vartheta_1,\vartheta_2,\phi_3)
  &:=&
  -\overline{n}_{\rm g}\e^{-\i(\varphi_1+\varphi_2)}
  \int\d^2\theta\,
  \Omega\big(|\vec{\theta}_1-\vec{\theta}|,
  |\vec{\theta}_2-\vec{\theta}|,
  |\vec{\theta}_1-\vec{\theta}_2|\big)
  \overline{\gamma}_{\rm h}(\vec{\theta}_3-\vec{\theta})
  \\\nonumber
  &=&-\overline{n}_{\rm g}\e^{-\i(\varphi_1+\varphi_2)}
  \int\d^2\theta\,
  \Omega\big(|\vec{\theta}_{13}+\vec{\theta}|,
  |\vec{\theta}_{23}+\vec{\theta}|,\vartheta_3\big)
  \,\overline{\gamma}_{\rm h}(\vec{\theta})\\
  &=&\nonumber
  \overline{n}_{\rm g}\int\d\theta\,\theta\,\d\varphi\,
  \Omega\big(\Upsilon(\vartheta_1,\theta,\varphi-\varphi_1),
  \Upsilon(\vartheta_2,\theta,\varphi-\varphi_2),\vartheta_3\big)
  \,\e^{-\i(\varphi_1+\varphi_2-2\varphi)}
  \overline{\gamma}_{\rm t,h}(\theta)\\
  &=&\nonumber
  \overline{n}_{\rm g}\int\d\theta\,\theta\,\d\varphi\,
  \Omega\big(\Upsilon(\vartheta_1,\theta,\varphi+\phi_3),\Upsilon(\vartheta_2,\theta,\varphi),\vartheta_3\big)
  \,\e^{+2\i(\phi_3+\varphi)}
  \overline{\gamma}_{\rm t,h}(\theta)\\\nonumber\\
  &=&  
  \label{eq:g3h}
  \overline{n}_{\rm g}\e^{+2\i\phi_3}
  \!\!\int_0^{\infty}\!\!\!\!\!\d\theta\,\theta\,\overline{\gamma}_{\rm t,h}(\theta)
  \!\!\int_0^{2\pi}\!\!\!\!\!\!\d\varphi\cos{(2\varphi)}
  \,\Omega\big(\Upsilon(\vartheta_1,\theta,\varphi+\phi_3),
  \Upsilon(\vartheta_2,\theta,\varphi),\vartheta_3\big)\;, 
\end{eqnarray}
for which we have introduced the auxiliary function
\begin{equation}
  \label{eq:psi2}
  \Upsilon(\theta_1,\theta_2,\phi):=
  \sqrt{\theta_1^2+\theta_2^2
    +2\theta_1\theta_2\cos{\phi}}\;.
\end{equation}
The transformations in \Ref{eq:g3h} use $\phi_3=\varphi_2-\varphi_1$
and a change of the integral variables
$\varphi\mapsto\varphi+\varphi_2$ and
$\vec{\theta}\mapsto\vec{\theta}+\vec{\theta}_3$. The last step
utilises that $\Omega(\ldots)$ has a vanishing imaginary part.

\begin{figure*}
  \begin{center}
    \epsfig{file=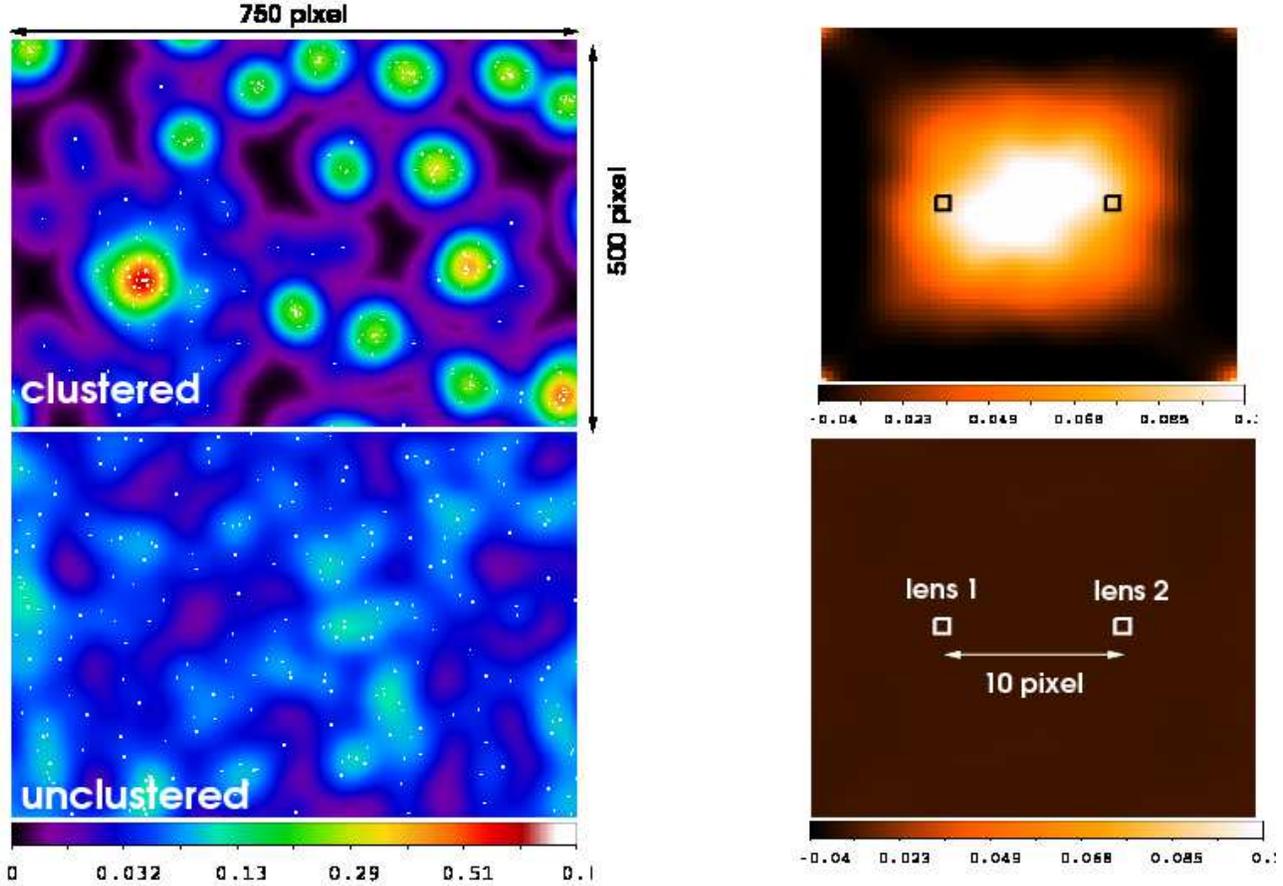,width=170mm,angle=0}
  \end{center}
  \caption{\label{fig:mockdata} \emph{Right panels:} Excess mass
    around lens pairs with fixed separation; squares indicate the lens
    positions inside the maps. \emph{Left panels:} Excerpts of the
    underlying ILM mock data: Lenses, shown as little dots, are either
    clustered (top left) or randomly distributed on the sky (bottom
    left). For simplicity, every lens has the same individual matter
    halo with a Gaussian lensing convergence profile (r.m.s. size is
    20 pixel) sticked to it.  The clustered lens haloes produces the
    joint matter halo of galaxy clusters in this model. The intensity
    scale in the left panels depicts the combined lensing convergence
    of all lenses; this is probed as shear by a sample of source
    galaxies. Note that the angular scale or the shear amplitude are
    of no particular interest here. The bottom right panel is the
    actual measurement of the bottom left scenario with the colour
    scale of the top right panel.}
\end{figure*}

\subsection{Interpretation}

In the context of $\cal G$, we can define a \emph{excess mass} map in
the following way. The function $\cal G$ can, for lens pairs of fixed
separation $\vartheta_3$, be mapped as excess shear field at position
$\vec{\theta}_3$ with Cartesian shear value
\begin{equation}
  \gamma_{\rm
    c}\left(\vec{\theta}_3|\vec{\theta}_1,\vec{\theta}_2\right)=
  -\frac{\vec{\theta}_{13}\vec{\theta}_{23}}
    {|\vec{\theta}_{13}||\vec{\theta}_{23}|}\,
  {\cal G}\left(\theta_{13},\theta_{23},\phi_3\right)\;,
\end{equation}
where $\phi_3$ is defined as angle spanned by $\vec{\theta}_{23}$ and
$\vec{\theta}_{13}$. In this map, we fix the lens positions at
$\vec{\theta}_1=+\vartheta_3/2$ and $\vec{\theta}_2=-\vartheta_3/2$ on
the $x$-axis. This shear map can be converted into a convergence map
\cite[e.g.][]{kas93}, as for example done in
\citet{2008A&A...479..655S}. The excess mass from the two-halo term of
the ILM is just the halo mass about each lens at $\vec{\theta}_1$ and
$\vec{\theta}_2$, weighed with the clustering strength
$\omega(\vartheta_3)$ of the pair. This is exactly the mass we would
anticipate around a pair of lenses, after one has subtracted the mass
around unclustered pairs (Eq. \ref{eq:g2h} with
$\omega(\vartheta_3)\equiv1$) and if one ignored the effect of third
haloes. The excess mass stemming from a third halo, clustering around
the lens pair, is described by the three-halo term.

That the excess shear or mass originates from galaxy clusters can be
argued from the ILM (Fig. \ref{fig:mockdata}).  In the ILM, the excess
shear is expressed in terms of the $2^{\rm nd}$- and $3^{\rm
  rd}$-order lens clustering correlation functions on the sky. We have
no signal, if lenses are unclustered, i.e., \mbox{$\omega=\Omega=0$},
or simply if we have no galaxy clusters. Unclustered lenses with
statistically independent matter haloes cannot produce any excess
mass.  On the other hand, they still may generate a GGL signal
\Ref{eq:ggl} if \mbox{$\overline{\gamma}_{\rm h}\ne0$}. If lenses
cluster, i.e., \mbox{$\omega\ne0$} or \mbox{$\Omega\ne0$}, we will get
automatically non-vanishing contributions to $\cal G$. For lens-lens
separations comparable or smaller than the typical angular size of a
cluster, most lens pairs will inhabit clusters and contribute mostly
to the excess mass. Therefore, those pairs probe essentially the
matter environment of clusters, provided they are at similar
redshift. We do not expect relevant contribution to $\cal G$ from
pairs of lenses with distinct redshifts (apparent pairs),
though. Imagine a catalogue of lenses in which all galaxies are
clearly separated in radial distance. On the sky, these lenses are (a)
unclustered and (b) their matter environments are mutually
statistically independent owing to the large physical distances
between lenses. This exactly covers the aforementioned situation as
reflected in a ILM with vanishing $\omega$ and $\Omega$: $\cal G$ from
this lens catalogue vanishes. In comparison with a sample of galaxies
all at similar radial distance, a survey with radial spread in the
lens distribution will have a larger fraction of apparent pairs, a
reduced angular clustering of lenses and hence a overall suppressed
amplitude of $\cal G$. This suppression can be corrected for, if the
radial distribution of lenses is specified (Appendix \ref{sect:cal}).

The ILM is only an approximation for the clustering of galaxies and
matter since a change of the matter-to-light ratio with size of
structures is not possible. Every structure can only be a sum of
individual haloes with no correlation to each other. Contrary to GGL
this limitation is relevant for G3L, as discussed in the following. We
first notice that in comparison with GGL, the lens-lens-shear
correlator under ILM assumptions seemingly does not provide any
fundamentally new information about the lens-matter connection. To
clarify this, the clustering correlation functions $(\omega,\Omega)$
can be determined by the observable lens angular distribution without
exploiting the gravitational lensing effect. Then, by utilising
$\omega$ and the observed mean tangential shear
$\overline{\gamma}_{\rm t}$, the average halo shear profile
$\overline{\gamma}_{\rm h}$ can be constrained from Eq.\Ref{eq:ggl} so
that all essential ingredients for predicting $\cal G$,
Eq. \Ref{eq:gtilde}, are already fixed. In particular, $\cal G$
appears to be only sensitive to the average halo shear profile as GGL
is. Seen this way, G3L can at most complement constraints on the mean
shear profile $\overline{\gamma}_{\rm h}$. In a scenario more complex
than the ILM, however, this differs.  Imagine throwing in a few
completely different matter haloes into a ILM, hosting several
galaxies simultaneously, that cannot be described as sums of
individual haloes.  Conventional GGL would be unable to detect a
difference to a ILM scenario, as we can still use the clustering
correlation function $\omega$ and the GGL signal $\bar{\gamma}_{\rm
  t}$ with \Ref{eq:ggl} to define an average lens shear profile
$\bar{\gamma}_{\rm h}$.  Therefore the ILM is always, even when
falsely presumed, sufficient to consistently describe the GGL signal
and the lens clustering. On the other hand, when then combined with
$\cal G$, we would observe inconsistencies, as we fail to correctly
explain $\cal G$ with Eq. \Ref{eq:gtilde}. In fact, the new haloes
with more than one galaxy would produce a one-halo term (Appendix
\ref{sect:onehalo}), which is missing in the ILM description. From
this we conclude that $\cal G$, in combination with GGL, enables us to
detect whether a ILM sufficiently explains the data or whether a more
advanced description is required.

With the ILM as reference scenario we suggest to construct a test for
the applicability of the ILM with the excess signal $\Delta\cal G$
constructed as follows:
\begin{enumerate}
\item Use GGL, lens clustering statistics and numbers to obtain the
  ILM parameter set $(\overline{n}_{\rm
    g},\omega,\Omega,\bar{\gamma}_{\rm h})$ via Eq.  \Ref{eq:ggl};
\item Define the ILM excess signal by
  \begin{equation}
    \Delta{\cal G}(\vartheta_1,\vartheta_2,\phi_3):=
    {\cal G}(\vartheta_1,\vartheta_2,\phi_3)-
    {\cal G}_{\rm 2h}(\vartheta_1,\vartheta_2,\phi_3)-
    {\cal G}_{\rm 3h}(\vartheta_1,\vartheta_2,\phi_3)\;,
  \end{equation}
  where the last two terms on the r.h.s. are the two- and three-halo
  term, Eqs. \Ref{eq:g2h} and \Ref{eq:g3h}, from the ILM description.
\end{enumerate}
A vanishing $\Delta\cal G$ tests the validity of a ILM description for
the data or expresses the deviation from it. 

\section{Lens-shear-shear correlator}
\label{sect:gpm}

Here we predict a measurement for the second G3L correlator
$\wtilde{G}_\pm$, given through the correlation of two shears and one
lens number density
\begin{equation}
  \Ave{
    \gamma_{\rm c}(\vec{\theta}_1)
    \gamma^\pm_{\rm c}(\vec{\theta}_2)
    n_{\rm g}(\vec{\theta}_3)}=
  \bar{n}^{-1}_{\rm g}\e^{+2\i(\varphi_1\mp\varphi_2)}
  \widetilde{G}_\pm(\vartheta_1,\vartheta_2,\phi_3)
\;.
\end{equation}
The geometry of the correlator is depicted in the middle panel of
Fig. \ref{fig:gglsketch}.  Here and in the following equations, a
superscript ``$\pm$'' as in $\gamma_{\rm c}^\pm$ means $\gamma_{\rm
  c}$ for $\gamma_{\rm c}^-$ and $\gamma_{\rm c}^+$ for the complex
conjugate $\gamma_{\rm c}^\ast$.  This correlator measures the
shear-shear correlations as function of lens separation. As before
with $\cal G$, symmetries demand that the correlator depends only on
relative separations and angles given by the triangle defined by lens
and source positions. It contains an unconnected part that describes
the shear-shear correlations for randomly distributed lenses with no
correlation to the shear field, namely (SW05)
\begin{equation}
  \label{eq:gpmunconnected}
  \widetilde{G}_+^{\rm nc}(\vartheta_1,\vartheta_2,\phi_3):=
  \xi_+(\vartheta_3)\,\e^{+2\i\phi_3}~;~
  \widetilde{G}_-^{\rm nc}(\vartheta_1,\vartheta_2,\phi_3):=
  \xi_-(\vartheta_3)
  \left(\frac{\vartheta_2}{\vartheta_3}\e^{+\i\phi_3/2}-
    \frac{\vartheta_1}{\vartheta_3}\e^{-\i\phi_3/2}\right)^4\;;
\end{equation}
$\vartheta_3$ denotes the source-source separation.  As before, the
unconnected terms do not contribute to the aperture statistics, here
$\ave{{\cal N}M_{\rm ap}^2}$, and are thus of no particular interest
for G3L. Subtracting the unconnected terms leaves us with the relevant
excess shear-shear correlations about lenses, formally
(cf. Eq. \ref{eq:rotshear})
\begin{equation}
  G_\pm(\vartheta_1,\vartheta_2,\phi_3)=
  \Ave{\gamma(\vec{\theta}_1;\varphi_1)
    \gamma^\pm(\vec{\theta}_2;\varphi_2)
    \kappa_{\rm g}(\vec{\theta}_3)}
  =\wtilde{G}_\pm(\vartheta_1,\vartheta_2,\phi_3)-
  \wtilde{G}^{\rm nc}_\pm(\vartheta_1,\vartheta_2,\phi_3)\;.
\end{equation}

\subsection{Derivation}

The evaluation of $G_\pm$ for the ILM boils down to evaluating the
connected terms of the triple correlator
\begin{eqnarray}
  \label{eq:gpmcalc}
  \lefteqn{\Ave{
      \gamma_{\rm c}(\vec{\theta}_1)
      \gamma^\pm_{\rm c}(\vec{\theta}_2)
      n_{\rm g}(\vec{\theta}_3)}}\\
  &=&\nonumber
 \underbrace{\sum_{i}^{N_{\rm d}}
   \Ave{\delta_{\rm D}^{(2)}(\vec{\theta}_3-\vec{\theta}^{\rm h}_i)
   \gamma_{\rm h}(\vec{\theta}_1-\vec{\theta}^{\rm h}_i;\vec{\alpha}_i)
   \gamma^\pm_{\rm h}(\vec{\theta}_2-\vec{\theta}^{\rm
     h}_i;\vec{\alpha}_i)}_i}_{\rm one-halo}
 +
 \underbrace{\sum_{i\ne j \ne k}^{N_{\rm d}}
   \Ave{\delta_{\rm D}^{(2)}(\vec{\theta}_3-\vec{\theta}^{\rm h}_i)
   \gamma_{\rm h}(\vec{\theta}_1-\vec{\theta}^{\rm h}_j;\vec{\alpha}_j)
   \gamma^\pm_{\rm
     h}(\vec{\theta}_2-\vec{\theta}^{\rm
     h}_k;\vec{\alpha}_k)}_{i,j,k}}_{\rm three-halo}\\
 &+&\nonumber
 \underbrace{\sum_{i\ne j}^{N_{\rm d}}
   \Ave{\delta_{\rm D}^{(2)}(\vec{\theta}_3-\vec{\theta}^{\rm h}_i)
   \gamma_{\rm h}(\vec{\theta}_1-\vec{\theta}^{\rm h}_j;\vec{\alpha}_j)
   \gamma^\pm_{\rm
     h}(\vec{\theta}_2-\vec{\theta}^{\rm
     h}_i;\vec{\alpha}_i)}_{i,j}
 +
 \sum_{i\ne j}^{N_{\rm d}}
 \Ave{\delta_{\rm D}^{(2)}(\vec{\theta}_3-\vec{\theta}^{\rm h}_i)
   \gamma_{\rm h}(\vec{\theta}_1-\vec{\theta}^{\rm h}_i;\vec{\alpha}_i)
   \gamma^\pm_{\rm
     h}(\vec{\theta}_2-\vec{\theta}^{\rm
     h}_j;\vec{\alpha}_j)}_{i,j}}_{\rm two-halo-2}\\
 &+&\nonumber
 \underbrace{\sum_{i\ne j}^{N_{\rm d}}
   \Ave{\delta_{\rm D}^{(2)}(\vec{\theta}_3-\vec{\theta}^{\rm h}_j)
   \gamma_{\rm h}(\vec{\theta}_1-\vec{\theta}^{\rm h}_i;\vec{\alpha}_i)
   \gamma^\pm_{\rm
     h}(\vec{\theta}_2-\vec{\theta}^{\rm
     h}_i;\vec{\alpha}_i)}_{i,j}}_{\rm two-halo-1}\;,
\end{eqnarray}
which now contains a one-halo term, two-halo terms and a three-halo
term. We distinguish two categories of two-halo terms: In
``two-halo-1'', the two shear signals are associated with the same
halo, while in ``two-halo-2'' the shear signals originate from the
lens halo and a different neighbouring halo.  In analogy to the
calculations for the correlator $\cal G$, only terms associated with
the leading order clustering correlation functions in the halo
correlators $\ave{\ldots}_{i,j}$ (terms with $\omega$) and
$\ave{\ldots}_{i,j,k}$ (terms with $\Omega$) are of interest for the
connected terms; for $\ave{\ldots}_i$ all terms are connected. Going
through the averages step by step and collecting the connected terms,
yields as final result for
\mbox{$\vec{\theta}_1\ne\vec{\theta}_2\ne\vec{\theta}_3$}
\begin{equation}
  \label{eq:gpm}
  G_\pm(\vartheta_1,\vartheta_2,\phi_3)=
  G_\pm^{\rm 1h}(\vartheta_1,\vartheta_2,\phi_3)
  +
  G_\pm^{\rm 2h1}(\vartheta_1,\vartheta_2,\phi_3)+
  G_\pm^{\rm 2h2}(\vartheta_1,\vartheta_2,\phi_3)
  +
  G_\pm^{\rm 3h}(\vartheta_1,\vartheta_2,\phi_3)\;.
\end{equation}

We start with the three-halo term, which is after performing the
integral variable transformations
\mbox{$\vec{\theta}\mapsto\vec{\theta}+\vec{\theta}_3$} and
\mbox{$\vec{\theta}^\prime\mapsto\vec{\theta}^\prime+\vec{\theta}_3$}
\begin{eqnarray}
  G_\pm^{\rm 3h}(\vartheta_1,\vartheta_2,\phi_3)
    &:=&
    \overline{n}_{\rm g}^2\e^{-2\i(\varphi_1\mp\varphi_2)}
    \int
    \d^2\theta\,\d^2\theta^\prime  
    \Omega\big(|\vec{\theta}-\vec{\theta}_3|,
    |\vec{\theta}^\prime-\vec{\theta}_3|,
    |\vec{\theta}-\vec{\theta}^\prime|\big)\,
    \overline{\gamma}_{\rm h}(\vec{\theta}_1-\vec{\theta})
    \overline{\gamma}^\pm_{\rm h}(\vec{\theta}_2-\vec{\theta}^\prime)\\
    &=&\nonumber
    \overline{n}_{\rm g}^2\e^{-2\i(\varphi_1\mp\varphi_2)}
    \int
    \d^2\theta\,\d^2\theta^\prime  
    \Omega\big(
    |\vec{\theta}|,  
    |\vec{\theta}^\prime|,
  |\vec{\theta}^\prime-\vec{\theta}|\big)
  \,\overline{\gamma}_{\rm h}(\vec{\theta}_{13}+\vec{\theta})
  \overline{\gamma}^\pm_{\rm
    h}(\vec{\theta}_{23}+\vec{\theta}^\prime)\;.
\end{eqnarray}
To cast this into a form that no longer explicitly contains any
$\varphi_i$, we need to do a few more transformations. We first note
that for a shifted tangential shear one has
\begin{equation}
  \overline{\gamma}_{\rm h}(\vec{\theta}+\vec{\theta}^\prime)=
  -\frac{\vec{\theta}+\vec{\theta}^\prime}
  {(\vec{\theta}+\vec{\theta}^\prime)^\ast}
  \,\overline{\gamma}_{\rm t,h}(|\vec{\theta}+\vec{\theta}^\prime|)
  =
  -\e^{+2\i\varphi^\prime}\Delta\left(\frac{\theta}{\theta^\prime},\varphi-\varphi^\prime\right)
  \,\overline{\gamma}_{\rm t,h}\big(\Upsilon(\theta,\theta^\prime,\varphi-\varphi^\prime)\big)\;,
\end{equation}
where 
\begin{equation}
  \Delta(s,\phi):=\frac{s\e^{+\i\phi}+1}{s\e^{-\i\phi}+1}=
  \frac{1+2\,s\cos{\phi}+s^2\cos{(2\phi)}
  -\i 2s(1+s\cos{\phi})\sin{\phi}}{1+s^2+2s\cos{\phi}}~;~
  \Delta^\ast(s,\phi)= \Delta^{-1}(s,\phi)
\end{equation}
is an additional phase factor; $\varphi$ and $\varphi^\prime$ are the
polar angles of $\vec{\theta}$ and $\vec{\theta}^\prime$,
respectively; $\Upsilon$ is given by the previous Eq. \Ref{eq:psi2}.
Then this allows us to rewrite the previous equation for $G_\pm^{\rm
  3h}$ as
\begin{eqnarray}
  \nonumber
  \lefteqn{G_\pm^{\rm 3h}(\vartheta_1,\vartheta_2,\phi_3)}
  \\\nonumber
  &:=&\overline{n}_{\rm g}^2\,\cancel{\e^{-2\i(\varphi_1\mp\varphi_2)}}
  \int\d^2\theta\,\d^2\theta^\prime  
  \Omega\big(
  |\vec{\theta}|,  
  |\vec{\theta}^\prime|,
  |\vec{\theta}^\prime-\vec{\theta}|\big)
  \,\cancel{\e^{2\i\varphi_1}}
  \Delta\left(\frac{\theta}{\vartheta_1},\varphi-\varphi_1\right)
  \overline{\gamma}_{\rm t,h}(|\vec{\theta}_{13}+\vec{\theta}|)
  \,\cancel{\e^{\mp2\i\varphi_2}}
  \Delta^\pm\left(\frac{\theta^\prime}{\vartheta_2},\varphi^\prime-\varphi_2\right)
  \overline{\gamma}_{\rm
    t,h}(|\vec{\theta}_{23}+\vec{\theta}^\prime|)\\
  &=&\nonumber
  \overline{n}_{\rm g}^2
  \int\d\theta\,\theta\,\d\theta^\prime\theta^\prime\d\varphi\,\d\varphi^\prime
  \Delta\left(\frac{\theta}{\vartheta_1},\varphi-\varphi_1\right)
  \Delta^\pm\left(\frac{\theta}{\vartheta_2},\varphi^\prime-\varphi_2\right)
  \Omega\big(\theta,  
  \theta^\prime,\Upsilon(\theta,\theta^\prime,\varphi^\prime-\varphi)\big)
  \,\overline{\gamma}_{\rm
    t,h}\big(\Upsilon(\vartheta_1,\theta,\varphi-\varphi_1)\big)
  \,\overline{\gamma}_{\rm
    t,h}\big(\Upsilon(\vartheta_2,\theta^\prime,\varphi^\prime-\varphi_2)\big)\\\nonumber\\
  &=&
  \label{eq:gpm3h}
  \overline{n}_{\rm g}^2
  \int_0^\infty\!\!\d\theta\,\theta\int_0^\infty\!\!\d\theta^\prime\theta^\prime
  \int_0^{2\pi}\!\!\d\varphi\int_0^{2\pi}\!\!\d\varphi^\prime
  \Delta\left(\frac{\theta}{\vartheta_1},\varphi\right)
  \Delta^\pm\left(\frac{\theta}{\vartheta_2},\varphi^\prime\right)
  \Omega\big(\theta,\theta^\prime,
  \Upsilon(\theta,\theta^\prime,\varphi^\prime-\varphi+\phi_3)\big)
  \,\overline{\gamma}_{\rm
    t,h}\big(\Upsilon(\vartheta_1,\theta,\varphi)\big)
  \,\overline{\gamma}_{\rm
    t,h}\big(\Upsilon(\vartheta_2,\theta^\prime,\varphi^\prime)\big)\;.
\end{eqnarray}

The (connected) two-halo terms can be split into two sub-groups. One
group is insensitive to shape variations of the lens halo, as it only
contains the mean halo shear profile $\overline{\gamma}_{\rm h}$,
\begin{eqnarray}
  \nonumber
  G_\pm^{\rm 2h2}(\vartheta_1,\vartheta_2,\phi_3)
  &:=&
  \overline{n}_{\rm g}\e^{-2\i(\varphi_1\mp\varphi_2)}
  \int\d^2\theta\,
  \omega\big(|\vec{\theta}_3-\vec{\theta}|\big)
  \big(
  \overline{\gamma}_{\rm h}(\vec{\theta}_1-\vec{\theta}_3)
  \overline{\gamma}^\pm_{\rm h}(\vec{\theta}_2-\vec{\theta})+
  \overline{\gamma}_{\rm h}(\vec{\theta}_1-\vec{\theta})
  \overline{\gamma}^\pm_{\rm h}(\vec{\theta}_2-\vec{\theta}_3)
  \big)\\
  &=&\nonumber
  \overline{n}_{\rm g}\e^{-2\i(\varphi_1\mp\varphi_2)}
  \int\d^2\theta\,
  \omega(\theta)
  \big(
  \overline{\gamma}_{\rm h}(\vec{\theta}_{13})
  \overline{\gamma}^\pm_{\rm
    h}(\vec{\theta}_{23}+\vec{\theta})+
  \overline{\gamma}^\pm_{\rm h}(\vec{\theta}_{23})
  \overline{\gamma}_{\rm h}(\vec{\theta}_{13}+\vec{\theta})
  \big)\\
  &=&\nonumber
  \bar{n}_{\rm g}\overline{\gamma}_{\rm t,h}(\vartheta_1)
  \int_0^\infty\d\theta\,\theta
  \,\overline{\gamma}_{\rm t,h}(\theta)
  \int_0^{2\pi}\d\varphi
  \cos{(2\varphi)}\,\omega\big(\Upsilon(\vartheta_2,\theta,\varphi)\big)\\  
  &+&\nonumber
  \bar{n}_{\rm g}\overline{\gamma}_{\rm t,h}(\vartheta_2)
  \int_0^\infty\d\theta\,\theta
  \,\overline{\gamma}_{\rm t,h}(\theta)
  \int_0^{2\pi}\d\varphi
  \cos{(2\varphi)}\,\omega\big(\Upsilon(\vartheta_1,\theta,\varphi)\big)\\
  \nonumber\\
  &=&\label{eq:gpm2h2}
 \overline{\gamma}_{\rm t,h}(\vartheta_1)
  \big(\overline{\gamma}_{\rm t}(\vartheta_2)-\overline{\gamma}_{\rm
    t,h}(\vartheta_2)\big)+
  \overline{\gamma}_{\rm t,h}(\vartheta_2)
  \big(\overline{\gamma}_{\rm t}(\vartheta_1)-\overline{\gamma}_{\rm
    t,h}(\vartheta_1)\big)\;.
\end{eqnarray}
The last step exploits the two-halo term of our previous result
\Ref{eq:ggl} for GGL. The terms inside the brackets express the excess
tangential shear due to haloes clustering around the lens halo. The
steps in the calculation of $G_\pm^{2h2}$ are by and large identical
to the steps undertaken in Sect. \ref{sect:gglcalc}.

The remaining terms in \Ref{eq:gpm} are special, as they are indeed
sensitive to halo variations, which sets them clearly apart from all
aforementioned correlation functions. The (connected) one-halo term is
\begin{equation}
  G_\pm^{\rm 1h}(\vartheta_1,\vartheta_2,\phi_3)
  :=
  \e^{-2\i(\varphi_1\mp\varphi_2)}
  \Ave{
    \gamma^{}_{\rm h}(\vec{\theta}_{1}-\vec{\theta}_3;\vec{\alpha})
    \gamma^\pm_{\rm
      h}(\vec{\theta}_{2}-\vec{\theta}_3;\vec{\alpha})}_{\vec{\alpha}}
  =
  \e^{-2\i(\varphi_1\mp\varphi_2)}
  \Ave{
    \gamma^{}_{\rm h}(\vec{\theta}_{13};\vec{\alpha})
    \gamma^\pm_{\rm
      h}(\vec{\theta}_{23};\vec{\alpha})}_{\vec{\alpha}}\;.
\end{equation}
If we write the halo shear as sum of the mean shear profile and some
fluctuation $\delta\gamma_{\rm h}(\vec{\theta};\vec{\alpha})$ about
it, i.e.,
\begin{equation}
  \gamma_{\rm h}(\vec{\theta};\vec{\alpha})=
  \overline{\gamma}_{\rm h}(\vec{\theta})+
  \delta\gamma_{\rm h}(\vec{\theta};\vec{\alpha})\;,
\end{equation}
with $\ave{\delta\gamma_{\rm
    h}(\vec{\theta};\vec{\alpha})}_{\vec{\alpha}}=0$, then the
one-halo term becomes
\begin{eqnarray}
 \nonumber
  G_\pm^{\rm 1h}(\vartheta_1,\vartheta_2,\phi_3)&=&
  \e^{-2\i(\varphi_1\mp\varphi_2)}\left(
  \overline{\gamma}_{\rm h}(\vec{\theta}_{13})
  \overline{\gamma}^\pm_{\rm h}(\vec{\theta}_{23})+
  \Ave{
    \delta\gamma_{\rm h}(\vec{\theta}_{13};\vec{\alpha})
    \delta\gamma^\pm_{\rm h}(\vec{\theta}_{23};\vec{\alpha})}_{\vec{\alpha}}
  \right)\\
  &=&\nonumber
  \overline{\gamma}_{\rm t,h}(\vartheta_1)
  \overline{\gamma}_{\rm t,h}(\vartheta_2)+
  \e^{-2\i(\varphi_1\mp\varphi_2)}
  \Ave{
    \delta\gamma_{\rm h}(\vec{\theta}_{13};\vec{\alpha})
    \delta\gamma^\pm_{\rm
      h}(\vec{\theta}_{23};\vec{\alpha})}_{\vec{\alpha}}\\
  &=&\label{eq:gpm1h}
  \overline{\gamma}_{\rm t,h}(\vartheta_1)
  \overline{\gamma}_{\rm t,h}(\vartheta_2)+
  \Ave{
    \delta\gamma_{\rm t,h}(\vartheta_1,\varphi_1;\vec{\alpha})
    \delta\gamma^\pm_{\rm
      t,h}(\vartheta_2,\varphi_2;\vec{\alpha})}_{\vec{\alpha}}\;.
\end{eqnarray}
In the last equation, we employed
\begin{equation}
  \gamma_{\rm t,h}(\theta,\varphi;\vec{\alpha}):=
  -\e^{-2\i\varphi}\,\gamma_{\rm
    h}(\vec{\theta};\vec{\alpha})~;~
  \overline{\gamma}_{\rm t,h}(\theta)=
  \Ave{\gamma_{\rm t,h}(\theta,\varphi;\vec{\alpha})}_{\vec{\alpha}}\;,
\end{equation}
where $\varphi$ is the polar angle of $\vec{\theta}$; an equivalent
definition is employed for the fluctuations $\delta\gamma_{\rm t,h}$.
Owing to statistical isotropy of the shear field, the model halo shear
profile $\gamma_{\rm h}(\vec{\theta};\vec{\alpha})$ has to have a
random orientation. Therefore, the correlator in the previous equation
must be invariant with respect to any rotation $\phi$, or
\begin{eqnarray}
  \Ave{
    \delta\gamma_{\rm t,h}(\vartheta_1,\varphi_1;\vec{\alpha})
    \delta\gamma^\pm_{\rm
      t,h}(\vartheta_2,\varphi_2;\vec{\alpha})}_{\vec{\alpha}}
  &=&\nonumber
  \frac{1}{2\pi}\int_0^{2\pi}\d\phi\,
  \Ave{
    \delta\gamma_{\rm t,h}(\vartheta_1,\varphi_1+\phi;\vec{\alpha})
    \delta\gamma^\pm_{\rm
      t,h}(\vartheta_2,\varphi_2+\phi;\vec{\alpha})}_{\vec{\alpha}}\\
  &=&\nonumber
  \frac{1}{2\pi}\int_0^{2\pi}\d\phi\,
  \Ave{
    \delta\gamma_{\rm t,h}(\vartheta_1,\phi+\phi_3;\vec{\alpha})
    \delta\gamma^\pm_{\rm
      t,h}(\vartheta_2,\phi;\vec{\alpha})}_{\vec{\alpha}}\\
  &=:& \label{eq:essgpm}
  \overline{\gamma}_{\rm t,h}(\vartheta_1)
  \overline{\gamma}_{\rm t,h}(\vartheta_2)
  \,\delta G_\pm^{1h}(\vartheta_1,\vartheta_2,\phi_3)\;.
\end{eqnarray}
The remaining two-halo term in \Ref{eq:gpm} is similar to the one-halo
term, actually an integral over $G^{\rm 1h}_\pm$,
\begin{eqnarray}
  \nonumber
  G_\pm^{\rm 2h1}(\vartheta_1,\vartheta_2,\phi_3)
  &:=&
  \overline{n}_{\rm g}\e^{-2\i(\varphi_1\mp\varphi_2)}
  \int\d^2\theta\,\omega\big(|\vec{\theta}-\vec{\theta}_3|\big)
  \Ave{
    \gamma_{\rm h}(\vec{\theta}_{1}-\vec{\theta};\vec{\alpha})
    \gamma^\pm_{\rm
      h}(\vec{\theta}_{2}-\vec{\theta};\vec{\alpha})}_{\vec{\alpha}}\\
  &=&\nonumber
  \overline{n}_{\rm g}\e^{-2\i(\varphi_1\mp\varphi_2)}
  \int\d^2\theta\,
  \omega(\theta)\,
  \Ave{
    \gamma_{\rm h}(\vec{\theta}_{13}+\vec{\theta};\vec{\alpha})
    \gamma^\pm_{\rm
      h}(\vec{\theta}_{23}+\vec{\theta};\vec{\alpha})}_{\vec{\alpha}}\\
  &=&\nonumber
  \overline{n}_{\rm g}
  \,\cancel{\e^{-2\i(\varphi_1\mp\varphi_2)}}
  \int\d^2\theta\,\omega(\theta)
  \,\cancel{\e^{+2\i\varphi_1}}
  \cancel{\e^{\mp2\i\varphi_2}}
  \Delta\left(\frac{\theta}{\vartheta_1},\varphi-\varphi_1\right)
  \Delta^\pm\left(\frac{\theta}{\vartheta_2},\varphi-\varphi_2\right)
  G^{\rm 1h}_\pm\big(
  \Upsilon(\vartheta_1,\theta,\varphi-\varphi_1),
  \Upsilon(\vartheta_2,\theta,\varphi-\varphi_2),\nu
  \big)\\\nonumber\\
  &=&
  \label{eq:gpm2h1}
  \overline{n}_{\rm g}
  \int_0^\infty\!\!\d\theta\,\theta\int_0^{2\pi}\!\!\!\d\varphi
  \,\omega(\theta)
  \Delta\left(\frac{\theta}{\vartheta_1},\varphi+\phi_3\right)
  \Delta^\pm\left(\frac{\theta}{\vartheta_2},\varphi\right)
  \,G^{\rm 1h}_\pm\big(
  \Upsilon(\vartheta_1,\theta,\varphi+\phi_3),
  \Upsilon(\vartheta_2,\theta,\varphi),\mu \big)\;,
\end{eqnarray}
where the angle $\nu$ spanned by $\vec{\theta}_{23}+\vec{\theta}$ and
$\vec{\theta}_{13}+\vec{\theta}$ and the corresponding angle $\mu$ for a
$\vec{\theta}$ rotated by $\varphi_2$ are implicitly given by
\begin{equation}
  \label{eq:angleaux}
  \e^{\i \nu}=
  \frac{\vec{\theta}_{23}+\vec{\theta}}
  {\vec{\theta}_{13}+\vec{\theta}}
  \frac{|\vec{\theta}_{13}+\vec{\theta}|}
  {|\vec{\theta}_{23}+\vec{\theta}|}
  =
  \frac{\vartheta_2\e^{-\i(\varphi-\varphi_2)}+\theta}
  {\vartheta_1\e^{-\i(\varphi-\varphi_1)}+\theta}
 \,\frac{\Upsilon(\vartheta_1,\theta,\varphi-\varphi_1)}
 {\Upsilon(\vartheta_2,\theta,\varphi-\varphi_2)}~;~
  \e^{\i\mu}=
  \frac{\vartheta_2\e^{-\i\varphi}+\theta}
  {\vartheta_1\e^{-\i(\varphi+\phi_3)}+\theta}
  \,\frac{\Upsilon(\vartheta_1,\theta,\varphi+\phi_3)}
  {\Upsilon(\vartheta_2,\theta,\varphi)}\;.
\end{equation}

\subsection{Interpretation}

The resulting $G_\pm$ is the lowest-order galaxy-galaxy lensing
correlation function that, at least within the framework of the ILM,
is sensitive to variations among shear profiles of
haloes. Nevertheless, we also have a $G_\pm$ signal when all halo
shear profiles are identical, i.e., $\delta\gamma_{\rm t,h}=0$, which
is generated by the tangential shear of the lens halo and the excess
tangential shear of clustering neighbouring haloes. Only
\Ref{eq:gpm1h} and \Ref{eq:gpm2h1}, which both contain the correlator
$\delta G^{\rm 1h}$, are affected by a scatter in halo matter
profiles.  The one-halo term of $G_\pm$ remains unchanged, even if we
allow for general correlations between halo parameters $\vec{\alpha}$
of distinct haloes or for correlations between lens positions and
$\vec{\alpha}$, since a statistical independence of haloes has not
been used for this term. Therefore, we expect the behaviour of $G_\pm$
on small angular scales to be described generally by $G_\pm^{\rm 1h}$,
not just within the ILM.

In the simplest case that all halo shear
profiles are exactly identical, $\gamma_{\rm
  h}(\vec{\theta};\vec{\alpha})=\overline{\gamma}_{\rm
  h}(\vec{\theta})$, we find
\begin{equation}
  G^{\rm 1h}_\pm(\vartheta_1,\vartheta_2,\phi_3)=
  \overline{\gamma}_{\rm t,h}(\vartheta_1)
  \overline{\gamma}_{\rm t,h}(\vartheta_2)~;~
  \delta G^{\rm 1h}(\vartheta_1,\vartheta_2,\phi_3)=0\;,
\end{equation}
i.e., the one-halo term has no explicit dependence on the opening
angle $\phi_3$. Note that $G^{\rm 1h}_+$ and $G^{\rm 1h}_-$ are 
identical. As illustration of the impact of variance in halo shear
profiles, consider a singular isothermal ellipsoid (SIE) profile with
ellipticity $\epsilon_{\rm h}$ and random orientation $\phi$
\citep{2006MNRAS.370.1008M}
\begin{equation}
  \gamma_{\rm h}^{\rm sie}(\vec{\theta})\propto
  -\frac{\e^{+2\i\varphi}}{\theta}\left(
    1+\frac{\epsilon_{\rm h}}{2}\cos{(2\varphi+2\phi)}\right)\;,
\end{equation}
$\varphi$ is the polar angle of $\vec{\theta}$. The absolute amplitude
of the shear profile is not of interest here. The tangential shear
profile of the SIE is
\begin{equation}
  \gamma_{\rm t,h}^{\rm sie}(\theta,\varphi)\propto
  \frac{1}{\theta}\left(
    1+\frac{\epsilon_{\rm h}}{2}\cos{(2\varphi+2\phi)}\right)\;.
\end{equation}
In this case, we find by marginalising over all orientations $\phi$,
\begin{equation}
  G^{\rm 1h}_\pm(\vartheta_1,\vartheta_2,\phi_3)=
  \frac{1}{2\pi}\int_0^{2\pi}\d\phi\,
  \gamma_{\rm t,h}^{\rm sie}(\vartheta_1,\phi_3)
  \left[\gamma_{\rm t,h}^{\rm
      sie}(\vartheta_2,0)\right]^\pm
  \propto
  \overline{\gamma}_{\rm t,h}(\vartheta_1)
  \overline{\gamma}_{\rm t,h}(\vartheta_2)
  \left(1+\frac{\epsilon_{\rm h}^2}{8}\cos{(2\phi_3)}\right)~;~
  \overline{\gamma}_{\rm t,h}(\vartheta)\propto\frac{1}{\vartheta}\;,
\end{equation}
thus similar to the previous result but now with an additional
$\phi_3$-dependent term, or
\begin{equation}
  \delta G_\pm^{\rm 1h}(\vartheta_1,\vartheta_2,\phi_3)=
  \frac{\epsilon_{\rm h}^2}{8}\cos{(2\phi_3)}\;.
\end{equation}
The result becomes somewhat more complicated for general slopes
$\delta$ (Appendix \ref{sect:nonspherical}) and will reveal a
difference between $G^{\rm 1h}_+$ and $G^{\rm 1h}_-$ when $\delta\ne1$
(not SIE) and $\epsilon_{\rm h}\ne 0$ (elliptical). Moreover, the
correlator \Ref{eq:essgpm} will exhibit no $\phi_3$-dependence, if the
lens haloes are always axially symmetric, even though their radial
matter density profile or their mass may scatter as to be expected in
reality.  Therefore, we conclude that $G_\pm$ may in principle be used
to constrain the shape or, more specifically, the mean second-moment
of the projected halo matter density profiles. In addition to that,
fluctuations $\delta\gamma_{\rm t,h}$ in the halo shear profile due to
halo substructure also add to the variance dependent one- and two-halo
term of $G_\pm$. As with the foregoing $\cal G$, it may be useful to
define an excess $\Delta G_\pm$, obtained by subtracting off the
$G_\pm$-signal as anticipated from the ILM with parameters from lens
clustering and GGL. In the ILM regime, the excess signal $\Delta
G_\pm$ exactly vanishes, if there is no scatter in the (projected)
matter density profiles.

\section{Conclusions}
\label{sect:discuss}

In order to gain a better understanding of G3L, we conceived a toy
model, the ``isolated lens model'' (ILM), for the distribution of
galaxies and matter about galaxies. In this picture, ``isolated''
galaxies are surrounded by their own matter envelope
(halo). Variations in the halo matter density profile are explicitly
allowed, albeit statistically independent to variations of other
matter envelopes or to positions of other lenses. Consequently, the
matter environment of clusters is herein the superposition of
independent haloes produced by clustering galaxies. The average
independent matter halo is described by the mean tangential shear
around lenses and the clustering of the lenses (GGL),
Eq. \Ref{eq:ggl}.  The foregoing calculations evaluate what would be
measured by G3L (Eqs. \ref{eq:gtilde} and \ref{eq:gpm}) under the ILM
assumptions and discuss the results. Here we summarise our main
conclusions.

\subsection{Excess shear about lens pairs}

The lens-lens-shear correlation function $\cal G$ basically stacks the
shear field about clustered lens pairs as opposed to GGL, which stacks
the shear field about individual lenses.  ``Clustered lens pairs''
refers to the fact that the connected part of the lens-lens-shear
correlator $\cal G$ does not include the expected shear pattern around
pairs formed by randomly distributed lenses.\footnote{Strictly
  speaking, $\widetilde{\cal G}$ is not the stacked, average shear
  field about lens pairs but the average shear pattern \emph{times}
  the frequency of lens pairs at separation $\vartheta_3$ normalised
  by the same frequency for randomly distributed lenses, i.e., times
  the factor $1+\omega(\vartheta_3)$.}  ``Excess mass'' refers to the
convergence map that corresponds to the excess shear of lens pairs.
Our conclusions are:
\begin{itemize}

\item Unclustered lenses do not generate any $\cal G$ signal, although
  they may exhibit a GGL signal. $\cal G$ is a probe for the matter
  environment of clusters (or groups), probed by lens pairs inhabiting
  the cluster.

\item Apparent lens pairs formed by lenses well separated in redshift,
  overall diminish the signal and add noise. The signal-to-noise of
  $\cal G$ can thus probably be improved by exploiting lens redshift
  information and by giving more weight to lenses that are close in
  redshift. The susceptibility of the $\cal G$ amplitude as to the
  radial distribution of lenses can be \ch{normalised} (Appendix
  \ref{sect:cal}).

\item In the ILM, the excess mass constituents are the haloes about
  the clustered lens pairs (two-halo term; Eq. \ref{eq:g2h}) and a
  third halo clustering about the clustered lens pair (three-halo
  term; Eq. \ref{eq:g3h}). In a more elaborate model incorporating
  haloes hosting more than one galaxy at a time, we expect also
  one-halo terms adding to the excess mass (Appendix
  \ref{sect:onehalo}), especially at small scales.

\item If the ILM is a fair description, then $\cal G$ does not provide
  any new information on the galaxy-matter connection compared to GGL
  combined with $2^{\rm nd}$-order galaxy clustering. $\cal G$ can at
  most complement the information on the mean halo shear profile.

\item G3L in combination with GGL and lens clustering can probe
  whether the matter environments of clusters could be expanded as
  sums of independent haloes, as in the ILM. In particular, a sum of
  independent haloes would be unable to change the mass-to-light ratio
  compared to that of field galaxies.

\item Moreover, the ILM can be employed as reference to quantify the
  deviation $\Delta\cal G$ from the ILM picture in the real matter
  distribution around galaxies. This can be devised as practical test
  for a more advanced halo-model picture that naturally presumes the
  possibility of genuine joint matter haloes of galaxies that are
  fundamentally different to sums of matter haloes about isolated
  lenses. In analogy to $\cal G$, $\Delta\cal G$ can be visualised as
  mass map about lens pairs or as aperture statistics.

\end{itemize}

\subsection{Excess shear-shear correlations about lenses}

Unlike GGL and $\cal G$, which both stack shear about lenses, $G_\pm$
measures the excess shear-shear correlations relative to a lens
position. ``Excess'' means in this context shear-shear correlations in
contrast to randomly distributed lenses, which would have a vanishing
GGL or $\cal G$ signal. In this respect, $G_\pm$ more resembles the
shear-shear correlation function $\xi_\pm$ utilised in cosmic shear
studies \citep[e.g.][]{2006glsw.conf..269S} rather than GGL.  Our
conclusions are:
\begin{itemize}
  
\item In the ILM, or more generally in a full halo model picture, the
  excess shear-shear correlations have two basic contributors: (i) the
  host halo of a lens (one-halo term; Eq. \ref{eq:gpm1h}), and (ii)
  neighbouring haloes clustering about the lens (two-halo terms:
  Eqs. \ref{eq:gpm2h1} and \ref{eq:gpm2h2}; three-halo term:
  Eq. \ref{eq:gpm3h}).

\item $G_\pm$ is the lowest-order galaxy-galaxy lensing correlation
  function that is sensitive to variations in the (projected) density
  profiles of matter around lenses. Traditional GGL and $\cal G$ are
  only functions of the mean, stacked shear profiles.

\item In particular is $G_\pm$ sensitive to variations due to
  elliptical haloes with random orientations. Therefore, the
  correlator is principally sensitive to the shape of matter haloes
  and could be exploited as such to measure halo shapes without the
  need of luminous tracers of presumed alignment to the halo. An
  elliptical halo generates an extra $\phi_3$-modulation in $G_\pm$,
  where $\phi_3$ is the opening angle between the two lens-source
  directions. The one-halo term of $G_\pm$ does not exhibit a
  $\phi_3$-modulation for spherical haloes.

\item Halo substructure, i.e., random fluctuations about a smooth halo
  profile, also contribute to the one- and two-halo term of
  $G_\pm$. Therefore, $G_\pm$ is in principle also sensitive to halo
  substructure.

\item The one-halo term of $G_\pm$ is unchanged, if we generally allow
  for statistical dependences of the lens haloes within the framework
  of a general model. Therefore, the ellipticity and substructure
  effect will also be present to some extend, if we have strong
  deviations from the ILM assumptions.  However, it is unclear at this
  point how strong the effects are, even within the ILM, and what
  possible degeneracies are. We defer a thorough study of these
  effects to a future paper.

\end{itemize}

\section*{Acknowledgements}
This work has been supported by the Deutsche Forschungsgemeinschaft in
the framework of the Collaborative Research Center TR33 `The Dark
Universe'.  Patrick Simon also acknowledges supported by the European
DUEL Research-Training Network (MRTN-CT-2006-036133).

\bibliographystyle{aa}
\bibliography{understandingG3L}

\appendix

\section{Normalisation scheme}
\label{sect:cal}

Let 
\begin{equation}
  {\Sigma_{\rm crit}(\chi_{\rm s},\chi_{\rm d}):=
  \frac{c^2}{4\pi Ga(\chi_{\rm d})}\frac{f_{\rm K}(\chi_{\rm s})}{f_{\rm K}(\chi_{\rm
      s}-\chi_{\rm d})f_{\rm K}(\chi_{\rm d})}}~;~
  \rho_{\rm crit}:=\frac{3H_0^2}{8\pi G}
\end{equation}
be the critical surface matter density for lenses at comoving distance
$\chi_{\rm d}$ and sources at $\chi_{\rm s}$, and the critical density
of the Universe, respectively; $f_{\rm K}(\chi)$ denotes the angular
diameter distance. Using Limber's equation for projecting the 3D
bispectrum to the angular 2D bispectrum, SW05 showed for
\begin{equation}
  |k|^2:=k_1^2+k_2^2-2 k_1k_2\cos{\psi}~;~
  A^2:= k_1^2R_1^2+ k_2^2R_2^2-2 k_1 k_2R_1R_2\cos{(\psi_3-\psi)}~;~
  \e^{2\i\nu}:=\frac{1}{A^2}
  \left[2 k_1 k_2 R_1 R_2+(k_1 R_1)^2
    \e^{\i(\phi_3-\psi)}+(k_2 R_2)^2
    \e^{\i(\phi_3-\psi)}\right]
\end{equation}
that
\begin{equation}
  {\cal G}(\vartheta_1,\vartheta_2,\phi_3)=
  \Omega_{\rm m}\rho_{\rm crit}
  \int_0^{\chi_{\rm h}}\frac{\d\chi\d\chi^\prime 
    p_{\rm b}(\chi^\prime)p_{\rm f}(\chi)^2}{\Sigma_{\rm crit}(\chi^\prime,\chi)}\,
  \int\frac{\d k_1k_1\d k_2 k_2\d\Psi}{(2\pi)^3}
  \underbrace{\frac{\left(k_1\e^{-\i\psi/2}+k_2\e^{+\i\psi/2}\right)^2}
  {|k|^2}\e^{2\i\nu}J_2(A)}_{=:K(k_1,k_2,R_1,R_2,\psi)}
B_{\rm ggm}\big(k_1,k_2,\psi;\chi\big)\;;
\end{equation}
$p_{\rm f}(\chi)$ and $p_{\rm b}(\chi)$ are the radial lens and source
distribution; $B_{\rm ggm}(k_1,k_2,\psi;\chi)$ is the
galaxy-galaxy-matter bispectrum at comoving radial distance $\chi$;
the wave numbers $k_i$ are also in comoving units.  Note that we here,
inside the integral, transformed angular separations $\vartheta_i$ to
the projected comoving distance at lens plane distance $\chi$,
$R_i:=f_{\rm K}(\chi)\vartheta_i$. We recast this equation into
\begin{eqnarray}
  {\cal G}(\overline{R}_1,\overline{R}_2,\phi_3)&:=&
  \Omega_{\rm m}\rho_{\rm crit}
  \int_0^{\chi_{\rm h}}\frac{\d\chi\d\chi^\prime 
    p_{\rm b}(\chi^\prime)p_{\rm f}(\chi)^2}
  {\Sigma_{\rm crit}(\chi^\prime,\chi)}\times
  \int\frac{\d k_1k_1\d k_2 k_2\d\psi}{(2\pi)^3}
  K(k_1,k_2,R_1,R_2,\psi)
  \,B_{\rm ggm}\big(k_1,k_2,\psi;\chi\big)\\
  &=:&\nonumber
  {\cal G}_0\times\int\frac{\d k_1k_1\d k_2 k_2\d\psi}{(2\pi)^3}
  K(k_1,k_2,R_1,R_2,\psi)
  \,B_{\rm ggm}\big(k_1,k_2,\psi;\overline{\chi}_{\rm d}\big)
\end{eqnarray}
where $B_{\rm ggm}\big(k_1,k_2,\psi;\overline{\chi}_{\rm d}\big)$ is
the bispectrum at effective lens plane distance $\overline{\chi}_{\rm d}$
defined by
\begin{equation}
  \int_0^{\chi_{\rm h}}\frac{\d\chi\d\chi^\prime 
    p_{\rm b}(\chi^\prime)p_{\rm f}(\chi)^2}
  {\Sigma_{\rm crit}(\chi^\prime,\chi)}
  B_{\rm ggm}\big(k_1,k_2,\psi;\chi\big)=:
  B_{\rm ggm}\big(k_1,k_2,\psi;\overline{\chi}_{\rm d}\big)
  \int_0^{\chi_{\rm h}}\frac{\d\chi\d\chi^\prime 
    p_{\rm b}(\chi^\prime)p_{\rm f}(\chi)^2}
  {\Sigma_{\rm crit}(\chi^\prime,\chi)}=
  B_{\rm ggm}\big(k_1,k_2,\psi;\overline{\chi}_{\rm d}\big)\,{\cal G}_0
\end{equation}
and $\overline{R}_i=f_{\rm K}(\overline{\chi}_{\rm d})\vartheta_i$ is
the projected comoving distance of angular separation $\vartheta_i$ at
the effective lens plane distance. Therefore, normalising $\cal G$ by
${\cal G}_0$ corrects the correlator for the amplitude reduction due
to lens pairs with lenses at distinct redshifts, encoded in $p_{\rm
  f}(\chi)$, and removes lensing related quantities, yielding a
bispectrum at an effective distance $\overline{\chi}_{\rm d}$
projected onto the lens plane by kernel $K(\ldots)$.

Similarly, we find for the lens-shear-shear correlator
\begin{equation}
  G_\pm\big(\overline{R}_1,\overline{R}_2,\phi_3\big)=
  G_{\pm_0}\times
  \int\frac{\d k_1k_1\d k_2 k_2\d\psi}{(2\pi)^3}
  K_\pm\big(k_1R_1,k_2R_2,\psi\big)
  \,B_{\rm mmg}\big(k_1,k_2,\psi;\overline{\chi}_{\rm d}\big)
\end{equation}
with
\begin{equation}
  G_{\pm_0}:=
  \Omega_{\rm m}^2\rho_{\rm crit}^2
  \int_0^{\chi_{\rm h}}\frac{\d\chi\d\chi^\prime 
    p_{\rm b}(\chi^\prime)p_{\rm f}(\chi)}
  {\Sigma_{\rm crit}(\chi^\prime,\chi)^2}
  ~;~
  \int_0^{\chi_{\rm h}}\frac{\d\chi\d\chi^\prime 
    p_{\rm b}(\chi^\prime)p_{\rm f}(\chi)}
  {\Sigma_{\rm crit}(\chi^\prime,\chi)^2}
  B_{\rm mmg}\big(k_1,k_2,\psi;\chi\big)=:
  B_{\rm mmg}\big(k_1,k_2,\psi;\overline{\chi}_{\rm d}\big)
  \,\frac{G_{\pm_0}}{\Omega_{\rm m}^2\rho_{\rm crit}^2}
\end{equation}
and the integral kernels
\begin{equation}
  K_+\big(k_1R_1,k_2R_2,\psi\big):=
  \e^{-2\i(\psi_3-\varphi_2)}J_0(A)~;~
  K_-\big(k_1R_1,k_2R_2,\psi\big):=
  \e^{+4\i\nu}J_4(A)\;.
\end{equation}

\section{One-halo term for $\cal G$ in a more complex model}
\label{sect:onehalo}

The ILM does not have any one-halo terms for $\cal G$ since haloes are
hosting only one lens at a time. Matter haloes of clusters have to be
represented as sums of independent haloes. In a more realistic
situation beyond the ILM, we can imagine genuinely new haloes, hosting
more than one galaxy, that cannot be expanded as sum of independent
haloes. For simplicity, for this new type of haloes we here assume
round haloes with shear profile $\overline{\gamma}_{\rm
  h}(\vec{\theta})$; $\vec{\theta}=0$ is the centre of the halo. Every
halo hosts the same number of $N_{\rm g}$ galaxies. Galaxies belonging
to the $i$th halo are scattered throughout the halo with a separation
$\vec{\Delta\theta}_{ij}$ relative to the halo centre
$\vec{\theta}^{\rm h}_i$. Now the number density of galaxies on the
sky and the (relevant) shear field associated with the new halo type
are
\begin{equation}
  n_{\rm g}(\vec{\theta})=
  \sum_{i=1}^{N_{\rm h}}\sum_{j=1}^{N_{\rm g}}\delta_{\rm
    D}(\vec{\theta}-\vec{\theta}^{\rm h}_i-\vec{\Delta\theta}_{ij})~;~
  \gamma_{\rm c}(\vec{\theta})=
  \sum_{i=1}^{N_{\rm h}}\overline{\gamma}_{\rm h}(\vec{\theta}-\vec{\theta}^{\rm h}_i)\;.
\end{equation}
In the following, we will focus on the connected one-halo terms of the
lens-lens-shear correlator only,
\begin{eqnarray}
  \label{eq:onehalo1}
  \Ave{n_{\rm g}(\vec{\theta}_1)n_{\rm g}(\vec{\theta}_2)\gamma_{\rm
      c}(\vec{\theta}_3)}&=&
  \sum_{i=1}^{N_{\rm h}}\sum_{j=1}^{N_{\rm d}}
  \Ave{
    \delta_{\rm
      D}(\vec{\theta}_1-\vec{\theta}^{\rm h}_i-\vec{\Delta\theta}_{ij})
    \delta_{\rm D}(\vec{\theta}_2-\vec{\theta}^{\rm h}_i-\vec{\Delta\theta}_{ij})
    \overline{\gamma}_{\rm
      h}(\vec{\theta}_3-\vec{\theta}_i)}_{i,j}\\
  \label{eq:onehalo2}
  &+&
  \sum_{i=1}^{N_{\rm h}}\sum_{j\ne k=1}^{N_{\rm d}}
  \Ave{
    \delta_{\rm
      D}(\vec{\theta}_1-\vec{\theta}^{\rm h}_i-\vec{\Delta\theta}_{ij})
    \delta_{\rm D}(\vec{\theta}_2-\vec{\theta}^{\rm h}_i-\vec{\Delta\theta}_{ik})
    \overline{\gamma}_{\rm
      h}(\vec{\theta}_3-\vec{\theta}_i)}_{i,j,k}\\
  &+&\texttt{(2-halo~terms)}+\texttt{(3-halo~terms)}\;,
\end{eqnarray}
where we  have the ensemble averages
\begin{equation}
  \Ave{\ldots}_{i,j}:=
  \frac{1}{A}\int\d^2\theta_i\,\d^2\Delta\theta_{ij}\,
  P_\Delta(|\vec{\Delta\theta}_{ij}|)\big[\ldots\big]~~~;~~~
  \Ave{\ldots}_{i,j,k}:=
  \frac{1}{A}\int\d^2\theta_i\,\d^2\Delta\theta_{ij}\,\d^2\Delta\theta_{ik}\,
  P_\Delta(|\vec{\Delta\theta}_{ij}|,|\vec{\Delta\theta}_{ik}|,
  |\vec{\Delta\theta}_{ij}-\vec{\Delta\theta}_{ik}|)
  \big[\ldots\big]\;.
\end{equation}
By $P_\Delta(\Delta\theta)$ we denote the p.d.f. of a single relative
galaxy position inside a halo, whereas by
$P_\Delta(\Delta\theta,\Delta\theta^\prime,|\vec{\Delta\theta}-\vec{\Delta\theta}^\prime|)$
we denote the joint p.d.f. of two galaxy positions. Owing to isotropy,
the former can only be a function of the modulus of
$\vec{\Delta\theta}$, the latter only a function of the relative
separations $|\vec{\Delta\theta}|$, $|\vec{\Delta\theta}^\prime|$ and
$|\vec{\Delta\theta}-\vec{\Delta\theta}^\prime|$. The mean number
density of galaxies populating the new haloes is $N_{\rm h}N_{\rm
  g}/A$. If the new halo type is mixed with other haloes, such as ILM
haloes, the \emph{total} number density of galaxies, $\bar{n}_{\rm
  g}$, may be different. The one halo term is not affected by the
presence of the other haloes apart from $\bar{n}_{\rm g}$.

The sum \Ref{eq:onehalo1} vanishes for
$\vec{\theta}_1\ne\vec{\theta}_2$ so that the only relevant
contribution to the one-halo terms of $\cal G$ is, after exploiting
the Dirac delta functions,
\begin{eqnarray}
  \nonumber
  {\cal G}^{\rm 1h}(\vartheta_1,\vartheta_2,\phi_3)&=&
  -\e^{-\i(\varphi_1+\varphi_2)}\bar{n}_{\rm g}^{-2}N_{\rm h}N_{\rm g}(N_{\rm g}-1)A^{-1}
  \int\d^2\theta\,
  P_\Delta\big(|\vec{\theta}+\vec{\theta}_{12}|,
  |\vec{\theta}|,|\vec{\theta}_{12}|\big)
  \,\overline{\gamma}_{\rm h}(\vec{\theta}_{32}+\vec{\theta})\\
  &=&\nonumber
  -\e^{-\i(\varphi_1+\varphi_2)}\bar{n}_{\rm
    g}^{-1}f_{\rm h}N_{\rm g}(N_{\rm g}-1)
  \int\d^2\theta\,
  P_\Delta\big(|\vec{\theta}+\vec{\theta}_{13}|,
  |\vec{\theta}+\vec{\theta}_{23}|,\vartheta_3\big)
  \,\overline{\gamma}_{\rm h}(\vec{\theta})\\
  &=&\nonumber
  \bar{n}_{\rm g}^{-1}f_{\rm h}N_{\rm g}(N_{\rm g}-1)
  \int\d\theta\,\theta\d\varphi\,
  P_\Delta\big(|\vec{\theta}+\vec{\theta}_{13}|,
  |\vec{\theta}+\vec{\theta}_{23}|,\vartheta_3\big)
  \e^{\i(2\varphi-\varphi_1-\varphi_2)}
  \,\overline{\gamma}_{\rm t,h}(\theta)\\\nonumber\\
  &=&
  \bar{n}_{\rm g}^{-1}\,\e^{+2\i\phi_3}f_{\rm h}N_{\rm g}(N_{\rm g}-1)
  \!\!\!\int_0^\infty\d\theta\,\theta\,\overline{\gamma}_{\rm t,h}(\theta)
  \int_0^{2\pi}\d\varphi\,\cos{(2\varphi)}\,
  P_\Delta\big(
  \Upsilon(\vartheta_1,\theta,\varphi+\phi_3),
  \Upsilon(\vartheta_2,\theta,\varphi),\vartheta_3\big)\;.
\end{eqnarray}
We used \Ref{eq:psi2} for the definition of $\Upsilon$. By $f_{\rm h}:=N_{\rm
  h}/(A\bar{n}_{\rm g})$ we mean the ratio of new type haloes to the
total number of galaxies; as usual,
$\vartheta_3^2=\vartheta_1^2+\vartheta_2^2-2\vartheta_1\vartheta_2\cos{\phi_3}$
is the separation of the lenses. If we add more haloes with, say,
different shear profiles or numbers of galaxies $N_{\rm g}$, we will
obtain a sum of one-halo terms of the previous kind, all weighed with
(i) their halo fractions $f_{\rm h}$ and (ii) number of galaxy pairs
$N_{\rm g}(N_{\rm g}-1)$. ILM haloes have trivially \mbox{$N_{\rm
    g}=1$}, thus vanishing one-halo terms due to the absence of galaxy
pairs. Moreover, if galaxies are unclustered inside their host haloes,
\mbox{$P_\Delta\sim\rm\,const$}, one will also have a vanishing
one-halo term. When the host haloes are not clustered as well, there
will be no contribution to $\cal G$ at all.

\section{General elliptical matter haloes and $G_\pm$}
\label{sect:nonspherical}

We consider a projected halo matter density profile with constant
power law slope when averaged over annuli. The profiles have an
elliptical shape with ellipticity $\epsilon_{\rm
  h}=(a^2-b^2)/(a^2+b^2)$; $a, b$ are the sizes of the major and minor
axis, respectively. \citet{2006MNRAS.370.1008M} find for the shear
profile corresponding to this matter density profile
\begin{equation} 
  \label{eq:mandelbaum06}
  \gamma_{\rm
    h}(\vec{\theta};\vec{\alpha})=
  -\frac{A\delta\theta^{-\delta}}{2-\delta}{\rm
    e}^{2{\rm i}\varphi}
  (\gamma_{\rm t}+{\rm i}\gamma_\times)~;~
  \gamma_{\rm t}:=
  1+\frac{(\delta-2)(\delta^2-2\delta+4)}{\delta^2(\delta-4)}
  \frac{\epsilon_{\rm h}}{2}\cos{(2\varphi)}~;~
  \gamma_\times:=
  \frac{4(2-\delta)(1-\delta)}{\delta^2(\delta-4)}\frac{\epsilon_{\rm
      h}}{2}\sin{(2\varphi)}\;,
\end{equation}
where we denote by $\varphi$ the polar angle of $\vec{\theta}$ and by
$A$ the amplitude of the shear.  For simplicity, we work with the
assumption that all haloes have the same ellipticity, power law index
and amplitude.  As the orientation of a halo is a-priori not known, we
have to marginalise over the random orientation angle of the halo for
correlator \Ref{eq:essgpm},
\begin{eqnarray}
  \overline{\gamma_{\rm t,h}}(\theta,\varphi) &=&
  \frac{A\delta\,\theta^{-\delta}}{\delta-2}\;,\\
  \delta G^{\rm 1h}_-(\vartheta_1,\vartheta_2,\phi_3)
  &=&
  \frac{\epsilon_{\rm
      h}^2}{8}f_-(\delta)\cos{(2\phi_3)}\;,
  \\
  \delta G^{\rm 1h}_+(\vartheta_1,\vartheta_2,\phi_3)
  &=&
  \frac{\epsilon_{\rm
      h}^2}{8}f_+(\delta)\cos{(2\phi_3)}
  +{\rm i}\frac{\epsilon_{\rm h}^2}{8}g_+(\delta)\sin{(2\phi_3)}\;,
\end{eqnarray}
with the auxiliary functions
\begin{equation}
  f_-(x) := \frac{(x-2)^3(x+2)}{x^3(x-4)}~;~
  f_+(x) := \frac{(x^4-4x^3+28x^2-48x+32)(x-2)^2}{(x-4)^2x^4}~;~
  g_+(x) := \frac{(x-2)^2(x^3-3x^2+6x-4)}{(x-4)^2x^4}\;.
\end{equation}
One has $f_+(\delta)=f_-(\delta)=1$ and $g_+(\delta)=0$ for a SIE
($\delta=1$).

\end{document}